\documentclass[11pt,a4paper]{article}

\usepackage{amsmath,amssymb,mathrsfs}
\usepackage{subfigure}
\usepackage{url}

\usepackage{graphicx}
\usepackage{color}

\usepackage{cite}

\usepackage{slashed}

\numberwithin{equation}{section}

\allowdisplaybreaks[4]

\begin{document}

\renewcommand{\subfigcapskip}{5pt}

\renewcommand{\theequation}{\thesection.\arabic{equation}}

\def\thefootnote{\fnsymbol{footnote}}


\hfill{} DESY 20-135

\hfill{} KANAZAWA-20-06

\hfill{} TUM-HEP-1279-20

\vspace{1.0 truecm}

\begin{center}

{\textbf{\LARGE
Primordial gravitational waves in a minimal \\ \vspace{3 truemm}
model of particle physics and cosmology
}}

\bigskip

\vspace{0.5 truecm}

{\bf 
Andreas Ringwald$^1$\footnote{andreas.ringwald@desy.de},
Ken'ichi Saikawa$^2$\footnote{saikawa@hep.s.kanazawa-u.ac.jp} and 
Carlos Tamarit$^{3}$\footnote{carlos.tamarit@tum.de}
} \\[5mm]

\begin{tabular}{lc}
&\!\! {$^1$ \em Deutsches Elektronen-Synchrotron DESY,}\\
&{\em Notkestra\ss e 85, D-22607 Hamburg, Germany}\\[.4em]
&\!\! {$^2$ \em Institute for Theoretical Physics, Kanazawa University, }\\
&{\em Kakuma-machi, Kanazawa, Ishikawa 920-1192, Japan}\\[.4em]
&\!\! {$^3$ \em Physik-Department T70, Technische Universit\"at M\"unchen,}\\
&{\em James-Franck Stra\ss e 1, D-85748 Garching, Germany}\\[.4em]
\end{tabular}

\vspace{1.0 truecm}

{\bf Abstract}
\end{center}

\begin{quote}
\hspace{0.5cm}
In this paper we analyze the spectrum of the primordial gravitational waves (GWs) predicted in the Standard Model*Axion*Seesaw*Higgs portal inflation (SMASH) model, 
which was proposed as a minimal extension of the Standard Model that addresses five fundamental problems of particle physics and cosmology 
(inflation, baryon asymmetry, neutrino masses, strong CP problem, and dark matter) in one stroke.
The SMASH model has a unique prediction for the critical temperature of the second order Peccei-Quinn (PQ) phase transition 
$T_c \sim 10^8\,\mathrm{GeV}$ up to the uncertainty in the calculation of the axion dark matter abundance,
implying that there is a drastic change in the equation of state of the universe at that temperature. Such an event is imprinted on the spectrum of GWs 
originating from the primordial tensor fluctuations during inflation and entering the horizon at $T \sim T_c$, which corresponds to 
$f \sim 1\,\mathrm{Hz}$,
pointing to a best frequency range covered by future space-borne GW interferometers.
We give a precise estimation of the effective relativistic degrees of freedom across the PQ phase transition and use it to evaluate the spectrum of GWs observed today.
It is shown that the future high sensitivity GW experiment -- ultimate DECIGO -- can probe
the nontrivial feature resulting from the PQ phase transition in this model.
\end{quote}

\thispagestyle{empty}

\newpage

\tableofcontents

\renewcommand{\thepage}{\arabic{page}}
\renewcommand{\thefootnote}{\arabic{footnote}}
\setcounter{footnote}{0}

\section{Introduction}
\label{sec:Introduction}
\setcounter{equation}{0}

One of the most robust and model-independent predictions of inflationary cosmology is a stochastic background of 
primordial gravitational waves (GWs) \cite{Grishchuk:1974ny,Starobinsky:1979ty} which originates from tensor fluctuations. 
The power spectrum of tensor fluctuations, when modes with comoving wave number $k$ exit 
the horizon during inflation, is proportional to the square of the Hubble expansion rate $H=\dot a/a$ during inflation:
\begin{equation}
\Delta_t^2 (k) = \left.\frac{2}{\pi^2}\frac{H^2}{M_P^2}\right|_{k = a_{\rm inf} H_{\rm inf}}
\approx  3.4\times 10^{-12}
\left(\frac{H_{\rm inf}}{10^{13}\,\mathrm{GeV}}\right)^2,
\label{eq:tensor_power_spectrum_def} 
\end{equation}
where   
$M_P\simeq 2.44\times 10^{18}\,\mathrm{GeV}$ is the reduced Planck mass and $a_{\rm inf}=a(t_{\rm inf})$ is the cosmic scale factor 
during inflation. 
The amplitude of the primordial GWs is
quantified in terms of the 
fractional contribution per logarithmic frequency interval ($f=k/(2\pi a_0)$, with $a_0$ the present scale factor), to the energy in the present universe, 
\begin{equation}
\Omega_{\rm gw}(f) \equiv \frac{1}{\rho_{\rm crit}}\frac{d\rho_{\rm gw}(f)}{d\ln f}
= \mathcal{T}_0(f)\, \Delta^2_t (f)\,, 
\label{Omega_gw_def}
\end{equation}
where $\rho_{\rm gw}$ is the energy density of GWs and $\rho_{\rm crit}= 3H_0^2 M_P^2$ the 
critical energy density in terms of today's  (i.e. at time $t=t_0$) Hubble expansion rate $H_0=H(t_0)\approx 3.24\times 10^{-18}\,h$\,Hz.

Clearly, a measurement of the overall normalization of the primordial GW spectrum
would yield crucial information about the dynamics of 
inflation, in particular its energy scale, $E_{\rm inf}=(3 H_{\rm inf}^2 M_P^2)^{1/4}$.
Moreover, a precise measurement of the frequency dependence of $\Omega_{\rm gw}(f)$ would provide also information about
the post-inflationary expansion 
history of the universe. 
The latter is encoded in Eq. \eqref{Omega_gw_def} in the transfer function $\mathcal{T}_0(f)\equiv \mathcal{T}(t_0,f)$, which 
accounts for the evolution of GWs after their modes reenter the horizon after inflation. 
Notably, it is sensitive to the 
equation of state of the post-inflationary universe at horizon crossing \cite{Schwarz:1997gv,Seto:2003kc,Saikawa:2018rcs}. 
In fact, for frequencies corresponding to modes which reenter the horizon during the post-reheating radiation-dominated epoch, 
$\frac{a_{\rm inf} H_{\rm inf}}{2\pi a_0} > \frac{a_{\rm rh} H_{\rm rh}}{2\pi a_0} \gg f\gg  \frac{a_{\rm eq} H_{\rm eq}}{2\pi a_0} 
\approx 1.6\times 10^{-17}$\,Hz, 
it can be approximated by \cite{Saikawa:2018rcs}
\begin{equation}
\mathcal{T}_0(f)\, h^2 
\approx 
3.2\times 10^{-6} 
\left(\frac{g_{\ast s,0}}{3.91}\right)^{\frac{4}{3}}
\left(\frac{g_{*\rho}(T_{\rm hc}(f))}{g_{*s}(T_{\rm hc}(f))}\right)^{\frac{4}{3}} 
\left[g_{*\rho}(T_{\rm hc}(f))\right]^{-\frac{1}{3}} 
\,,
\label{T_gw_gs_PT}
\end{equation}
where 
$g_{*\rho}(T_{\rm hc}(f))$ and $g_{*s}(T_{\rm hc}(f))$ denote the effective number of degrees of freedom of the
energy and the entropy density in the hot plasma at the temperature $T_{\rm hc}(f)$ of horizon crossing, respectively,
and $g_{\ast s,0}$ is the value of $g_{\ast s}$ after the neutrino decoupling.
The horizon crossing temperature $T_{\rm hc}$ can be related to
a frequency determined by the Hubble rate at this temperature via   
\begin{equation}
f = \frac{a_{\rm hc} H_{\rm hc}}{2\pi a_0}  \approx 
1.2
\,\mathrm{Hz}
\left(\frac{g_{\ast s,0}}{3.91}\right)^{\frac{1}{3}}
\left[ \frac{g_{*\rho}(T_{\rm hc})}{g_{*s}(T_{\rm hc})}\right]^{\frac{1}{2}}
\left[ {g_{*s}(T_{\rm hc})}\right]^{\frac{1}{6}}
\left(\frac{T_{\rm hc}}{10^8\,\mathrm{GeV}}\right)\,.
\label{f_to_Thc}
\end{equation}
According to this estimate, the spectrum of primordial GWs from inflation 
is expected to be almost flat for a huge range of frequencies, 
$10^{-17}\,{\rm Hz} \ll f\ll 
10^2\,{\rm Hz}\, (T_{\rm rh}/10^{10}\,{\rm GeV})$, and of order 
\begin{equation}
\Omega_{\rm gw}\, h^2
\approx 1.1\times 10^{-17}
\left(\frac{g_{\ast s,0}}{3.91}\right)^{\frac{4}{3}}
\left(\frac{g_{*\rho}(T_{\rm hc}(f))}{g_{*s}(T_{\rm hc}(f))}\right)^{\frac{4}{3}} 
\left[ g_{*\rho}(T_{\rm hc}(f))\right]^{-\frac{1}{3}} 
\left(\frac{H_{\rm inf}}{10^{13}\,\mathrm{GeV}}\right)^2.
\label{Omega_gw_est}
\end{equation} 
It has features, such as dips and steps, at frequencies corresponding to temperatures at which the equation of state changes considerably. 
Notably, $e^+e^-$ annihilation, the QCD and the electroweak crossover, occurring at temperatures around an MeV, 
100\,MeV and 100\,GeV, respectively, are predicted to show their imprints in the spectrum of primordial GWs at frequencies 
around $10^{-11}$\,Hz, $10^{-9}$\,Hz, and $10^{-6}$\,Hz, respectively. 

Unfortunately, there are no GW detectors foreseen which would be sensitive enough to detect the primordial GWs from inflation in this frequency range. 
Furthermore, the white dwarf (WD) confusion noise is an obstacle for detecting primordial GWs at frequencies $f \lesssim 0.1$\,Hz. 
So far, the only possible instruments to detect directly the primordial gravitational background from standard inflationary models 
appear to be future space-borne GW interferometers, like the Big Bang Observer (BBO)~\cite{BBO_proposal,Crowder:2005nr,Corbin:2005ny,Harry:2006fi} or the Deci-hertz Interferometer Gravitational Wave Observatory 
(DECIGO)~\cite{Seto:2001qf,Kawamura:2006up}. 
In fact, it has been argued that the ultimate sensitivity of DECIGO may become comparable to $\Omega_{\rm gw} h^2\sim 10^{-20}$\cite{Seto:2001qf} at frequencies around 1 Hz -- allowing not only for a large statistics detection of the primordial GWs from inflation, but also for a detailed investigation of possible features arising from changes in the equation of state at temperatures of the order of $10^8$\,GeV. 

A particularly well-motivated extension of the Standard Model (SM) which predicts a sizable change in the equation of state at such temperatures was developed in Refs. \cite{Ballesteros:2016euj,Ballesteros:2016xej,Ballesteros:2019tvf}. This minimal extension of the SM -- dubbed Standard Model*Axion*Seesaw*Higgs portal inflation (SMASH) -- addresses five fundamental problems of particle physics and cosmology (the origin of neutrino masses, the strong CP problem, the nature of dark matter, the generation of the matter-antimatter asymmetry in the universe, and the nature of the inflation) in one stroke. 
Because of its constrained framework, the model provides definite predictions for various cosmological observables that can be probed by upcoming axion dark matter and cosmic microwave background (CMB) experiments. In particular, it predicts a lower bound on the Hubble expansion rate during inflation, $H_{\rm inf}\gtrsim 10^{13}$\,GeV, and a lower bound on the ratio of the power in tensor to scalar fluctuations,
$r\gtrsim 0.004$, which can be probed by next generation CMB experiments sensitive to the predicted primordial $B$-mode polarization patterns, such as CMB-S4 \cite{Abazajian:2016yjj}, LiteBIRD \cite{Matsumura:2013aja}, and the Simons Observatory \cite{Ade:2018sbj}. Furthermore, it predicts a reheating temperature around $10^{10}$\,GeV, 
below which the universe has the equation of state of a thermal relativistic plasma. At around $T_c\sim 10^8$\,GeV, a global 
Peccei-Quinn (PQ) symmetry gets broken by a second order phase transition below which a number of particles get massive, leading to a 
sizable change in the effective number of relativistic degrees of freedom, $g_{*\rho}$ and $g_{*s}$, which will show up in a break in the spectrum of the primordial GWs at around $1\,$Hz. 

It is the purpose of this paper to provide a precise prediction of the primordial GW spectrum expected in SMASH
and to confront it with the projected sensitivity of ultimate DECIGO. 
This is done as follows. In Sec.~\ref{sec:model}, we give a brief review of SMASH, focusing on its inflationary dynamics, 
its predictions for the primordial tensor modes, and the nature and the critical temperature of the PQ phase transition. 
Section~\ref{sec:eos} deals with the precise calculation of the equation of state in SMASH. 
In Sec.~\ref{sec:GW}, we use the results from the previous sections to derive the primordial GW spectrum
and compare it with the projected experimental sensitivities. 
We discuss our results and conclude in Sec.~\ref{sec:conclusions}.

\section{The SMASH model}
\label{sec:model}
\setcounter{equation}{0}

In the SMASH model \cite{Ballesteros:2016euj,Ballesteros:2016xej,Ballesteros:2019tvf}, the SM is extended by adding a new complex scalar field $\sigma$ (the PQ field),
three singlet neutrinos $N_i$, with $i=1,2,3$, and a vector-like quark $Q$, all charged under a global $U(1)_{\rm PQ}$ symmetry. 
The scalar potential in SMASH has the general form 
\begin{equation}
\label{scalar_potential}
V(H,\sigma )= \lambda_H \left( H^\dagger H - \frac{v^2}{2}\right)^2
+\lambda_\sigma \left( |\sigma |^2 - \frac{v_{\sigma}^2}{2}\right)^2+
2\lambda_{H\sigma} \left( H^\dagger H - \frac{v^2}{2}\right) \left( |\sigma |^2 - \frac{v_{\sigma}^2}{2}\right)\,, 
\end{equation}
with $H$ the Higgs doublet, and $\lambda_H, \lambda_\sigma >0$,  
$\lambda_{H\sigma}^2 <  \lambda_H \lambda_\sigma$, such that the electroweak and PQ symmetry are broken by the vacuum expectation
values (VEVs) 
\begin{equation}
\langle H^\dagger H\rangle = v^2/2, \hspace{6ex}
\langle |\sigma |^2\rangle=v_{\sigma}^2/2\,,
\end{equation}
where $v_\sigma\gg v=246$\,GeV. The global PQ charges of the fermions and the hypercharge of the vector quark $Q$ can be chosen such that the only allowed  interactions of the exotic fermions $N_i,Q$ are 
\begin{align}\label{eq:LYuk}\begin{aligned}
 {\cal L}\supset &-\Bigg[F_{ij}\bar{ N}_j P_L L_i\epsilon H+\frac{1}{2}Y_{ij}\sigma \bar N_i P_L  N_j 
+y\, \sigma \bar Q P_L Q+\,{y_{Q_d}}_{i}\sigma\bar{D}_iP_L Q +h.c.\Bigg]\,.
\end{aligned}\end{align}
Here we used a four-component notation, with the $N_i$ represented by Majorana spinors, and $L_i,D_i$ denoting the Dirac spinors associated with the leptons and down quarks of the $i$th generation.
The axion arises as a Goldstone boson associated with the spontaneous breaking of the PQ symmetry \cite{Peccei:1977hh,Weinberg:1977ma,Wilczek:1977pj}. It 
can be the main constituent of dark matter if its decay constant $f_a \sim 10^{11}\,\mathrm{GeV}$ \cite{Preskill:1982cy,Abbott:1982af,Dine:1982ah},
where in SMASH we have $f_a = v_{\sigma}$.
This new energy scale also provides large Majorana masses for heavy neutrinos, which explains the smallness of the masses of the active neutrinos 
through the seesaw mechanism~\cite{Minkowski:1977sc,GellMann:1980vs,Yanagida:1979as,Mohapatra:1979ia} and leads to the generation of baryon asymmetry of the universe via the thermal leptogenesis scenario \cite{Fukugita:1986hr}.

In this framework, inflation can arise from the dynamics of the Higgs and PQ field in the presence of non-minimal couplings to the Ricci 
scalar $R$,
\begin{equation}
  \label{Lmain}
  S\supset - \int d^4x\sqrt{- g}\,\left[
     \frac{M^2}{2}  + \xi_H\, H^\dagger H+\xi_\sigma\, \sigma^* \sigma  
  \right] R
  \,,
\end{equation}
where the mass scale $M$ is related to the actual reduced Planck mass by  
\begin{align}
\label{eq:MMP}
M^2_P=M^2+\xi_H v^2+\xi_\sigma v^2_\sigma.
\end{align}
The non-minimal couplings  stretch the scalar potential in the Einstein frame, making it convex and asymptotically flat at large field values. In order to avoid problems with perturbative unitarity, it is required that $1\gtrsim \xi_\sigma\gg \xi_H\geq 0$. 
Furthermore, the requirement of a viable reheating demands that $\lambda_{H\sigma}<0$, in order that slow-roll inflation happens along the {line} $h/\rho=\sqrt{-\lambda_{H\sigma}/\lambda_H}$, where $\rho=\sqrt{2}\,|\sigma|$ is the modulus of the PQ field and $h$ the neutral component of the Higgs doublet in the unitary gauge. Inflation can then be described {in the Einstein frame} by a single canonically normalized field $\chi$ with potential
\begin{equation}
\label{genpotential}
\tilde V(\chi) = \frac{1}{4} \tilde \lambda_\sigma
\rho(\chi)^4\left(1+\xi_{\sigma}\frac{\rho(\chi)^2}{M_P^2}\right)^{-2}\,, \quad 
\tilde \lambda_\sigma \equiv  \lambda_\sigma \left( 1-\frac{\lambda_{H\sigma}^2}{\lambda_\sigma\lambda_H} \right)
\,.
\end{equation} 
The field $\chi$ is the solution of $\Omega^2\,d\chi/d\rho\simeq (b\,\Omega^2+6\,\xi_\sigma^2\,\rho^2/M_P^2)^{1/2}$, with $\Omega\simeq 1+\xi_\sigma\,\rho^2/M_P^2$ being the Weyl transformation into the Einstein frame and 
 $b=1+|\lambda_{H\sigma}/\lambda_H|$. Vacuum stability requires a small value of 
$|\lambda_{H\sigma}|\lesssim 10^{-6}$ and consequently $b\approx 1$. 

%
\begin{figure}[t]
\begin{center}
\includegraphics[width=5.5cm]{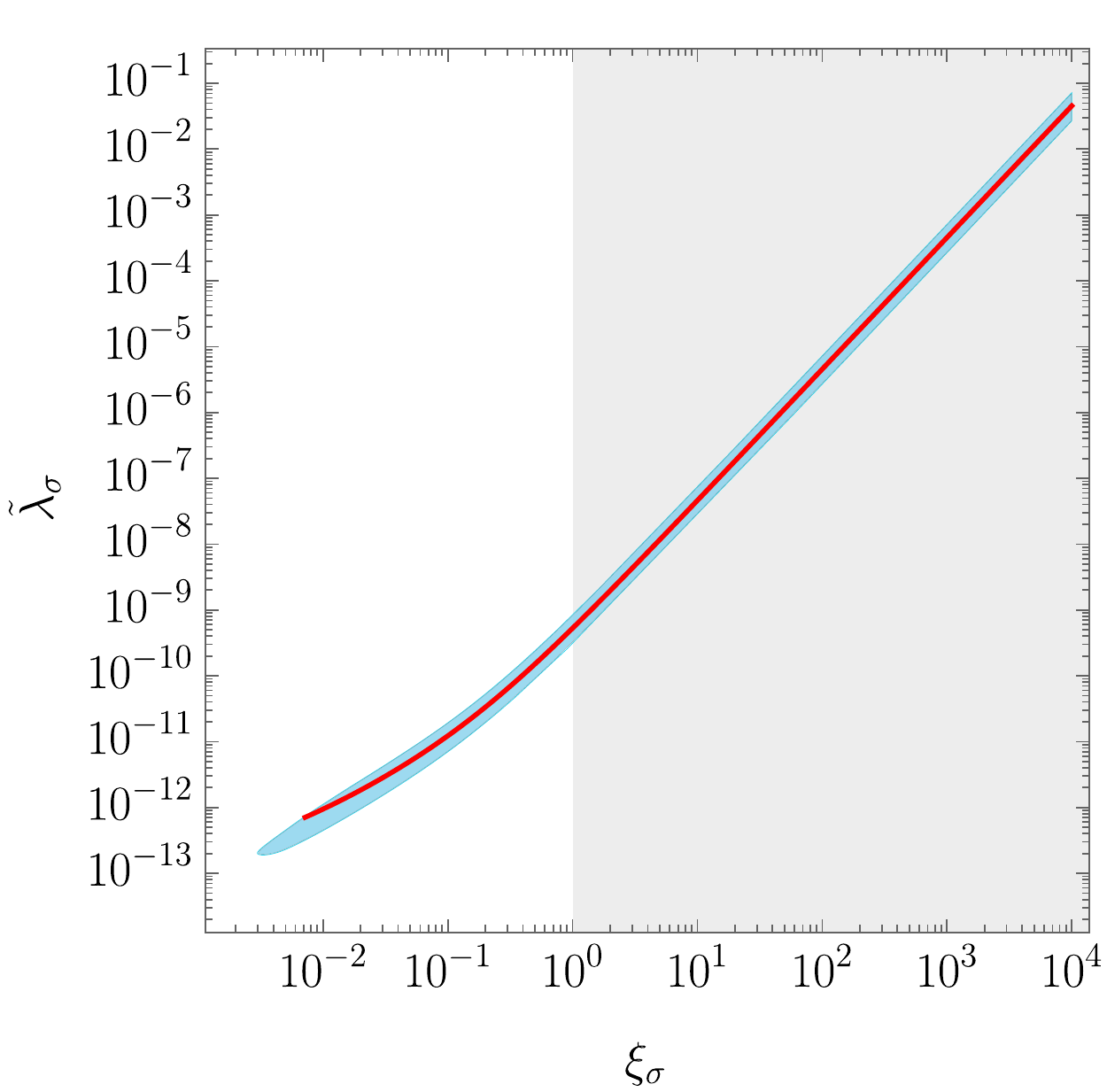} 
\includegraphics[width=5.5cm]{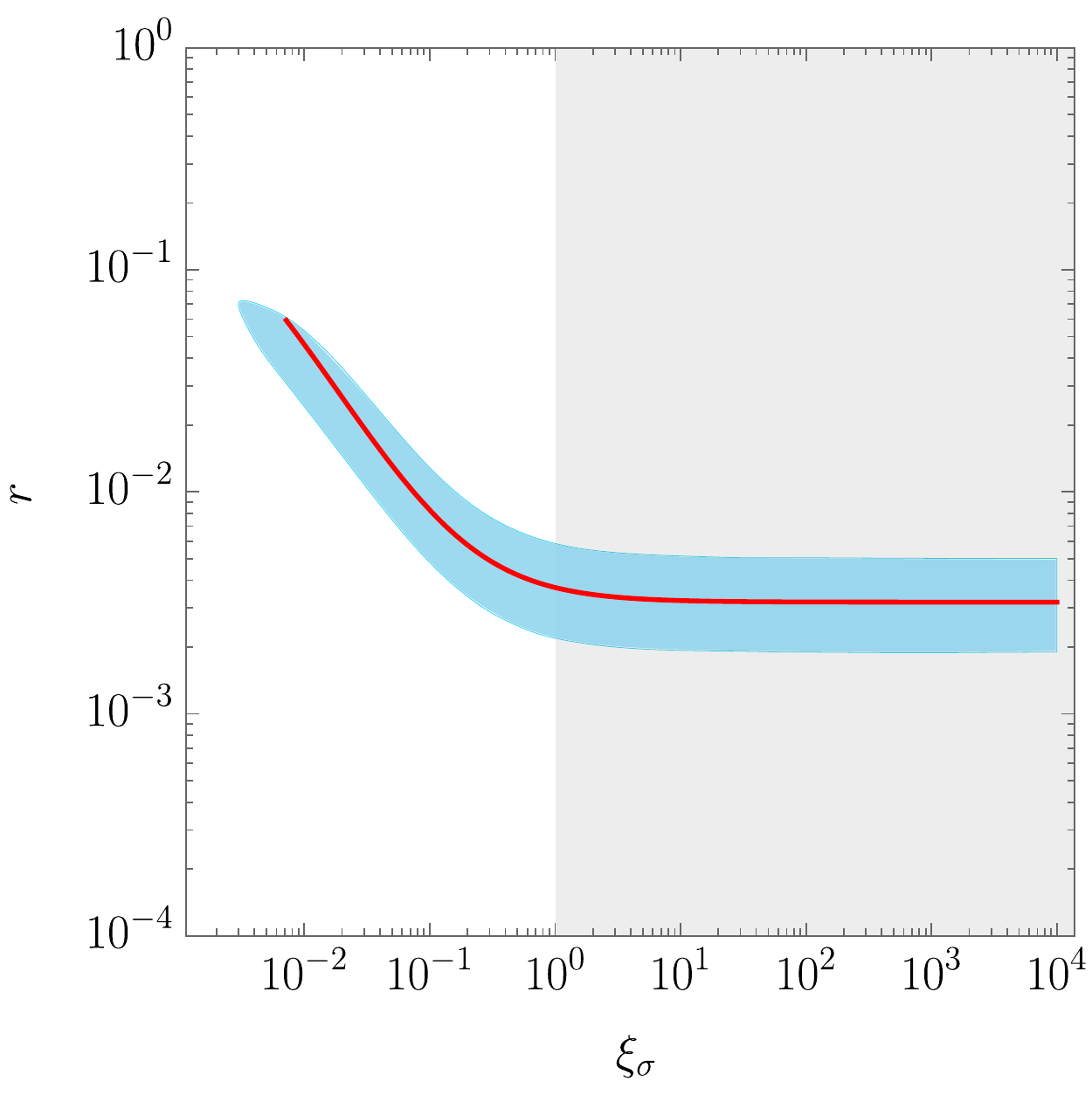}
\includegraphics[width=5.5cm]{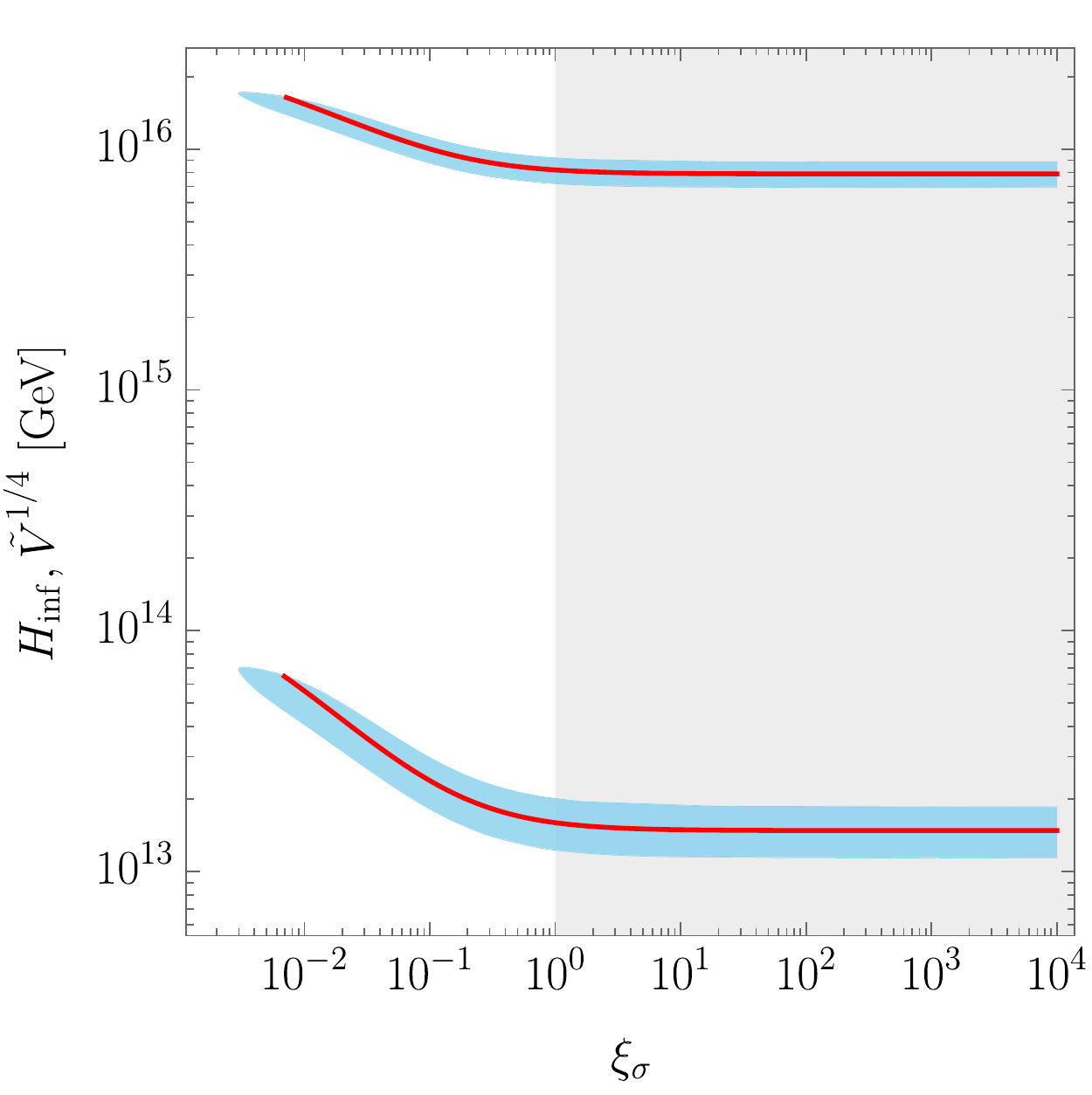}
\caption{95\% C.L. contours for the parameters of the non-minimally coupled potential \eqref{genpotential} giving inflation as constrained by Planck 2018 data at the pivot scale $0.002$ Mpc$^{-1}$ \cite{Aghanim:2018eyx,Akrami:2018odb}. Shown are: the value of the effective inflationary coupling (left), the predicted tensor-to-scalar ratio (middle), the Hubble rate during inflation, $H_{\rm inf}$ (right, lower band), and energy scale during inflation, $\tilde V^{1/4}$ (right, upper band), as a function of the non-minimal coupling parameter $\xi_\sigma$. The thicker red line corresponds to the predictions when accounting for the fact that reheating
in SMASH leads to radiation domination immediately after inflation. The shaded regions for $\xi_\sigma>1$ indicate, approximately, the region where the predictivity of inflation is threatened by the breakdown of perturbative unitarity.
}
\label{fig:smash_pot_par}
\end{center}
\end{figure}
%

The primordial scalar and tensor power spectra, $\Delta^2_s(k)$ and $\Delta^2_t(k)$, can be computed in the slow-roll approximation from the (potential) slow-roll parameters,
\begin{equation}
\epsilon=M_P^{2}(\tilde V'/\tilde V)\, , \quad 
\eta=M_P^2 \tilde V''/\tilde V\, , \quad
\zeta=M_P^4 \tilde V' \tilde V'''/\tilde V^2 \, , 
\end{equation}
where the primes denote derivatives with respect to $\chi$. 
Parametrizing the spectra as  
\begin{align}
\Delta^2_s(k)=A_s\left(k/k_*\right)^{n_s-1+{1/2\,\alpha}\log (k/k_*)+\cdots}\,,\quad 
\Delta^2_t(k)= A_t\left(k/k_*\right)^{n_t+\cdots}\,,
\end{align}
where all parameters are evaluated at some fiducial scale $k_*$, one obtains at leading order in the slow-roll expansion 
for the scalar spectral index $n_s$ and its running $\alpha \equiv dn_s/d\ln k$,
\begin{equation}
n_s \simeq 1-6\epsilon+2\eta \, , \quad \alpha \simeq  -2\zeta +16\epsilon\,\eta-24 \epsilon^2\,,
\end{equation}
and for the amplitude of scalar and tensor perturbations, $A_s$ and $A_t$,  
\begin{equation}
A_s \simeq  \frac{1}{24 \pi^2 \epsilon}\frac{\tilde V}{M_P^4} \, , \quad 
A_t \simeq \frac{2}{3 \pi^2}\frac{\tilde V}{M_P^4} \, , 
\end{equation}
evaluated at the field value corresponding to the time when the scale $k_*$ exits the horizon.  
Therefore, the tensor-to-scalar ratio is given by
\begin{equation}
r\equiv \frac{\Delta^2_t}{\Delta^2_s}\simeq \frac{A_t}{A_s}\simeq 16 \epsilon\,,
\end{equation}
while the tensor spectral index $n_t$ is entirely determined by $r$ through the consistency relation
\begin{equation}
n_t \simeq -\frac{r}{8}\,.
\label{eq:n_t_consistency_relation}
\end{equation}
Moreover, the number of e-folds from some initial field value $\rho_{\rm inf}$ until the end of inflation at $\rho_{\rm end}$ can also be approximately computed analytically: 
\begin{equation}
\label{eq:efoldssimple}
N(\chi_{\rm inf},\chi_{\rm end})\simeq 
\frac{1}{M_P}\int_{\chi_{\rm end}}^{\chi_{\rm inf}}\frac{d\chi}{\sqrt{2\epsilon}}=
\frac{b+6\xi_\sigma}{8 {M_P^2}}\left(\rho_{\rm inf}^2-\rho_{\rm end}^2\right)-\frac{3}{4}\ln\left(\frac{M_P^2+\xi_\sigma \rho_{\rm inf}^2}{M_P^2+\xi_\sigma \rho_{\rm end}^2}\right)\,.
\end{equation}
Actually, $N(\chi_{\rm inf},\chi_{\rm end})$ can be obtained exactly by solving the dynamics of the inflaton as a function of the number of e-folds itself, which is given by equation \cite{Ballesteros:2014yva}:
\begin{align} 
\label{eq:efolds}
\frac{d^2\chi}{dN^2}+3\,\frac{d\chi}{dN}-\frac{1}{2M_P^2}\left(\frac{d\chi}{dN}\right)^3+\left(3M_P-\frac{1}{2M_P}\left(\frac{d\chi}{d N}\right)^{2}\right)\sqrt{2\epsilon}=0\,,
\end{align}
and using the condition $\epsilon_H\equiv-\dot{{H}}/{H}^2=1$ to determine the value of $\chi$ at the end of inflation. 

\begin{figure}[h]
\centering
\includegraphics[width=0.9\linewidth]{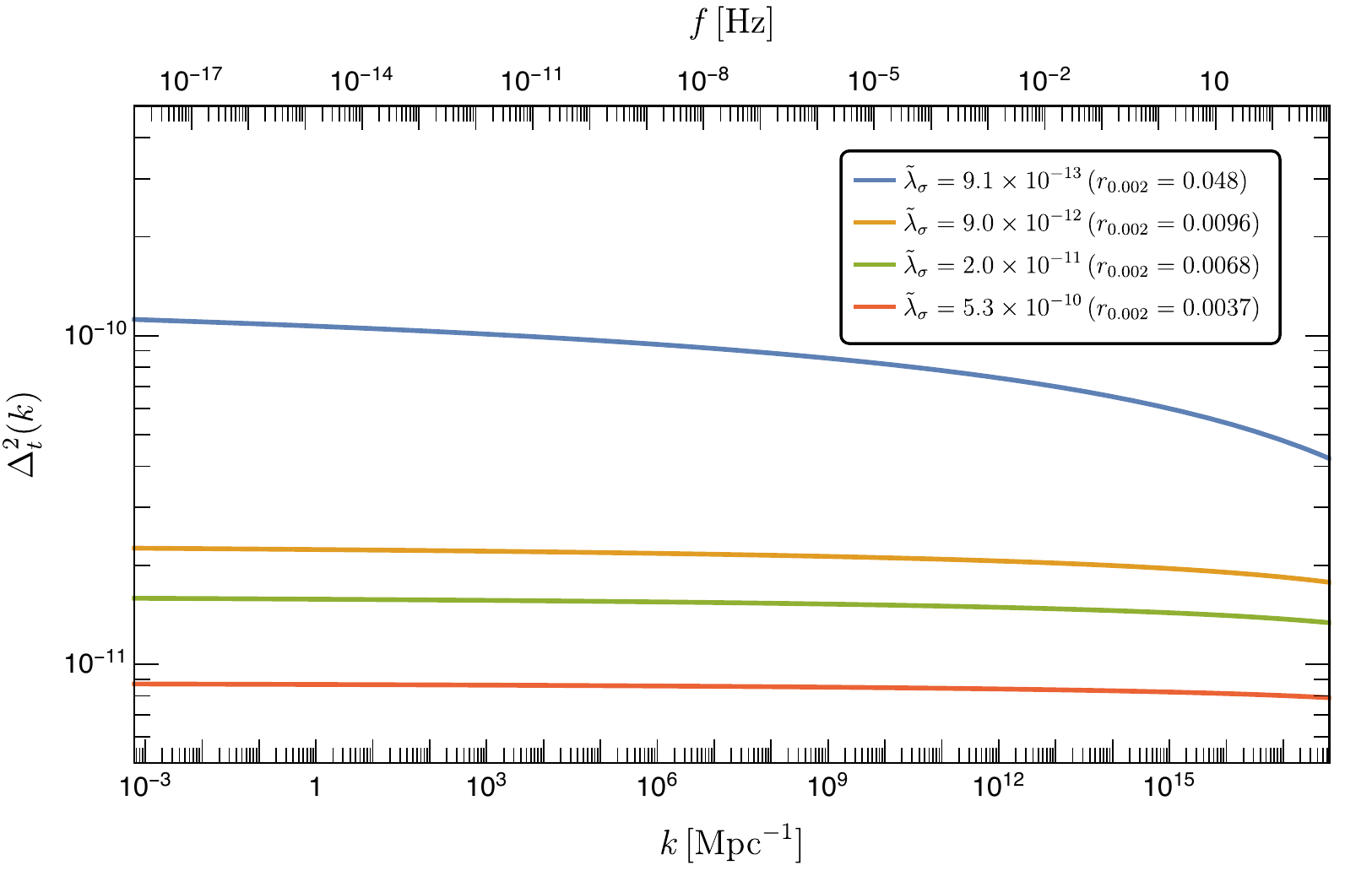}
\caption{Tensor power spectrum predicted in SMASH, for four different  values of the effective inflationary
coupling $\tilde{\lambda}_\sigma$.  
}
\label{fig:primordial_spectrum}
\end{figure}

Figure \ref{fig:smash_pot_par} 
shows parameters of the 
non-minimally coupled potential \eqref{genpotential} giving inflation as constrained by the CMB observations at Planck 2018
in combination with lensing, baryon acoustic oscillation (BAO), and BICEP-Keck Array 2015 (BK15) data~\cite{Aghanim:2018eyx,Akrami:2018odb},
\begin{eqnarray}
A_s&=&(2.105\pm 0.030)\times 10^{-9}\quad (68\,\%\,\text{C.L., TT,TE,EE+lowE+lensing+BAO}),
\\
n_s&=&0.9665 \pm 0.0038\quad (68\,\%\,\text{C.L., TT,TE,EE+lowE+lensing+BAO}),
\\
r_{0.002}&<& 0.058 \quad (95\,\%\,\text{C.L., TT,TE,EE+lowE+lensing+BK15+BAO}),
\end{eqnarray}
where the constraints on $A_s$ and $n_s$ are obtained at the pivot scale $k_* = 0.05$ Mpc$^{-1}$, and
that on $r$ is obtained at $k_* = 0.002$ Mpc$^{-1}$ (indicated by the subscript).
The corresponding tensor power spectrum predicted in SMASH is shown in Fig. \ref{fig:primordial_spectrum}, for four different values of the effective inflationary
coupling $\tilde{\lambda}_\sigma$. This will be used later in the determination of the spectrum of primordial GWs. 

Slow-roll inflation ends for $\rho\sim \mathcal{O}(M_P)$, when the inflaton field starts undergoing Hubble-damped oscillations in a quartic potential. The time-averaged stress-energy tensor for these oscillations has the equation of state of a radiation fluid. Hence, radiation domination starts right after inflation, and lasts through the phase of reheating in which the oscillating fields trigger the production of SM particles and the energy of the inflaton is transferred into the SM plasma. 
This post-inflationary history in a radiation-domination era  fixes the relation between the scales of the matter perturbations we observe in the universe today and the size of the primordial fluctuations which gave rise to them, when they outgrew the Hubble horizon and became frozen until their later horizon re-entry.  
This allows to compute the number of e-folds between horizon crossing and the end of inflation in a simple way. 
For a mode with comoving momentum $k$, we have  
\begin{equation}
\label{eq:Reheatquart}
 N_e(k)\simeq \log\frac{a_{\rm eq}\, {H}_{\rm eq}}{a_0\, {H}_0}-\frac{1}{4}\log\frac{3\,{H}_{\rm eq}^2}{M_P^2}-\log\frac{k}{a_0\,{H}_0}+\frac{1}{2}\log\frac{V_k}{M_P^4}+\frac{1}{4}\log\frac{M_P^4}{V_{\rm end}}\,,
\end{equation}
where $V_k$ and $V_{\rm end}$  denote, respectively, the energy density at the time of the mode's horizon crossing and at the end of inflation. 
By equating this expression with the result of integrating \eqref{eq:efolds} (or the simpler, but less accurate expression \eqref{eq:efoldssimple}) and using $A_s$ to fit $\tilde\lambda_\sigma$ we can compute the value of the inflaton field in the Jordan frame when a given scale $k$ exited the horizon (for a given $\xi_\sigma$) and give a definite prediction for $n_s$ and $r$.  
Using this expression one gets the thick red lines of Fig.~\ref{fig:smash_pot_par}. 

The Higgs component of the inflaton in SMASH guarantees efficient reheating by the production of SM gauge bosons.
The reheating temperature is predicted to be around {$T_{\rm rh}\sim  10^{10}$\,GeV.
Such a temperature ensures a thermal restoration of the PQ symmetry and its spontaneous breakdown later in the expansion 
history of the universe. 
The PQ phase transition in SMASH turns out to be second order. Its 
critical temperature can be estimated as~\cite{Ballesteros:2016xej}
\begin{equation}
T_c \simeq \frac{2\sqrt{6\lambda_{\sigma}}v_\sigma}{\sqrt{8(\lambda_{\sigma}+\lambda_{H\sigma})+\sum_iY_{ii}^2 + 6y^2}},
\label{eq:T_c_analytical}
\end{equation}
where $Y_{ii}$ and $y$ are the Yukawa couplings of $N_i$ and $Q$, respectively (see Eq.~\eqref{eq:LYuk}).
Taking into account that the parameters determining the critical temperature are constrained by the requirements of 
\begin{itemize}
\setlength{\labelsep}{0.4cm}
\item successful inflation compatible with the Planck constraint $r < 0.058$ and perturbative unitarity $\xi_{\sigma}<1$, which occurs for $5\times 10^{-13}\lesssim \tilde\lambda_{\sigma} \lesssim 5\times 10^{-10}$, 
\item vacuum stability, which requires $|\lambda_{H\sigma}|\lesssim Y_{ii}^2$ and $Y_{ii} \sim y \sim \lambda_{\sigma}^{1/4}$,
\item cold dark matter constituted by axions, which requires $10^{10}\,\mathrm{GeV} \lesssim v_\sigma \lesssim 2.2\times 10^{11}\,\mathrm{GeV}$,\footnote{
The value of $f_a$ (or $v_{\sigma}$) accounting for the observed cold dark matter abundance is subjected to the uncertainty in the calculation of the relic axion abundance.
Here we take the result of Ref.~\cite{Gorghetto:2020qws} as a lower limit and that of Ref.~\cite{Klaer:2017ond} as an upper limit. See Sec.~\ref{sec:conclusions} for more discussion.}
\end{itemize}
the critical temperature of the PQ phase transition is predicted to be in the range  
\begin{align}
T_c \sim \lambda_{\sigma}^{1/4}\,v_\sigma \sim \mathcal{O}(10^7\text{--} 10^9)\,\mathrm{GeV}.
\label{eq:T_c_estimate}
\end{align}
Around the critical temperature, there is a change in the equation of state of the primordial plasma, which will be worked out precisely 
in the next section.

\section{Equation of state for the Peccei-Quinn phase transition}
\label{sec:eos}
\setcounter{equation}{0}

In this section we review the computation of the effective numbers of relativistic degrees of freedom, $g_{\ast \rho}(T)$ and $g_{\ast s}(T)$, which affect the evolution of tensor perturbations after horizon crossing and thus leave an imprint in the observable power spectrum, as in Eq.~\eqref{Omega_gw_est}. The former quantities are related to the equation of state of the thermal plasma. They are linked to the total energy and entropy densities  as
\begin{align}\label{eq:rhosT}
 \rho=\frac{\pi^2}{30}g_{\ast\rho}T^4, \qquad s=\frac{2\pi^2}{45}g_{\ast s}T^3.
\end{align}
The first principle of thermodynamics and the Maxwell relation derived from the enthalpy function allow to relate the former quantities to the system's pressure:
\begin{align}
 \rho = T\frac{\partial p}{\partial T}-p, \qquad s=\frac{\partial p}{\partial T}.
\end{align}
In turn, for a system in equilibrium at constant temperature and volume and  in the absence of chemical potentials, the pressure can be related to minus the Helmholtz free-energy per unit volume, which itself is related to the partition function of the system at finite temperature. But the latter is also connected to the effective potential:
\begin{align}
 Z=e^{-\beta F}=e^{\beta V p}=e^{-\beta V V_{\rm eff}(T)}\Rightarrow p=-V_{\rm eff}(T),
\end{align}
where $V$ is the three volume.
As the system is assumed to be in equilibrium, i.e. in the configuration that minimizes the free-energy, the potential $V_{\rm eff}(T)$ is understood to be evaluated at its minimum. Thus, to compute $g_{\ast \rho}$ and $g_{\ast s}$, one just needs to know the values of the effective potential at its minimum as a function of temperature:
\begin{align}\label{eq:gs_computation}\begin{aligned}
 g_{\ast \rho}=&\,\frac{30}{\pi^2 T^4}\left(V_{\rm eff,min}(T)-T\frac{\partial V_{\rm eff,min}(T)}{\partial T}\right),\\
 g_{\ast s}=&\,-\frac{45}{2\pi^2 T^3} \frac{\partial V_{\rm eff,min}(T)}{\partial T}.
\end{aligned}\end{align}

In our calculations, we construct the effective potential in the Landau gauge (with gauge-fixing parameter $\xi=0$) including the one-loop Coleman-Weinberg potential, the one-loop thermal contributions, as well as higher-order QCD corrections up to three-loop order. As we focus on high temperatures at which the Higgs is stabilized at zero, we can consider a potential depending only on $\sigma$:
\begin{align}
 V_{\rm eff}(\sigma,T)=V(H=0,\sigma)+V^{\rm CW}(\sigma,T)+V^T(\sigma,T)+V^{\rm QCD}(T).
\label{eq:V_eff_full}
\end{align}
The tree-level piece $V(H=0,\sigma)$ follows from Eq.~\eqref{scalar_potential}, while $V^{\rm CW}(\sigma,T)$ designates the Coleman-Weinberg piece, $V^T(\sigma,T)$ is the thermal potential, and $V^{\rm QCD}(T)$ accounts for QCD contributions to the pressure. The temperature dependence in $V^{\rm CW}(\sigma,T)$ arises from the use of an improved daisy resummation in which the effective masses for bosonic fields appearing in 
$V^{\rm CW}$ (and  $V^T$) are substituted by appropriate temperature-dependent masses that incorporate the one-loop thermal corrections to the corresponding self-energies at zero momentum. Such a resummation is necessary because thermal corrections to the self-energies can dominate over their zero-temperature counterparts, such that diagrams that differ by insertions of thermal corrections to the propagators are all of the same order and need to be resummed. To leading order in momenta, the corrections to the propagators can be understood as thermal contributions to the masses. The resummed diagrams are known as ``daisy diagrams'',   and in the usual daisy resummation scheme, one simply incorporates the leading order thermal corrections to the bosonic masses in a high-temperature expansion, which go as $T^2$. However, as the high-temperature expansion is insensitive to mass thresholds, the usual daisy resummation is not compatible with decoupling, so that it cannot be used to follow the changes of $g_{\ast \rho}$ and $g_{\ast s}$ across a phase transition, which are partly due to the decoupling of particles from the plasma as the temperature drops below their mass. Hence it becomes crucial to implement a resummation that is compatible with thermal decoupling. In order to do so, we use an improved daisy resummation, which resums the self-energies at zero momentum but keeps the full temperature dependence. 

For computing $V^{\rm CW}(\sigma,T)$ we use the Landau gauge result in the $\overline{\rm MS}$ scheme,
\begin{align}
 V^{\rm CW}(\sigma,T)=\frac{1}{64\pi^2}&\left[\sum_V m^4_V(\sigma,T)\left(\log\frac{m^2_V(\sigma,T)}{\mu^2}-\frac{5}{6}\right)+\sum_S m^4_S(\sigma,T)\left(\log\frac{m^2_S(\sigma,T)}{\mu^2}-\frac{3}{2}\right)
 \right.\\
 \nonumber&\left.-\sum_F m^4_F(\sigma,T)\left(\log\frac{m^2_F(\sigma,T)}{\mu^2}-\frac{3}{2}\right)\right],
\end{align}
where $V, S, F$ denote massive gauge bosons, real scalars, and Weyl fermions, respectively, and where the sum over massive vectors includes a sum over the three polarizations that propagate in the Landau gauge, while the sum over fermion states also goes over the two spin/helicity states of each Weyl fermion.  $\mu$ designates the renormalization scale, while $m^2_{V/S/F}(\sigma,T)$ are the field-dependent masses in the background of $\sigma$, including the thermal corrections from the improved daisy resummation detailed in Appendix \ref{app:Daisy}. Note that massless bosons and ghosts do not contribute to the Coleman-Weinberg potential. 

The one-loop thermal potential $V^T$ is given by
\begin{align}
 \label{eq:VT}
  V^T=\frac{T^4}{2\pi^2}\left[\sum_B J_B\left(\frac{m^2_B(\sigma,T)}{T^2}\right)-\sum_F J_F\left(\frac{m^2_F(\sigma,T)}{T^2}\right)-\sum_GJ_B\left(0\right)\right],
 \end{align}
 where $B$ denotes all bosonic fields (massive and massless gauge bosons, counting three polarizations, and real scalars), $F$ denotes all Weyl fermions (counting 2 helicity states), and $G$ denotes complex ghosts---one for every generator of every gauge group---which are massless in the Landau gauge. Note that, in contrast to $V^{\rm CW}$, the massless fields contribute, which is crucial to get the correct physical values of $g_{\ast \rho}$ and $g_{\ast s}$. Despite ghosts being complex, there is no extra factor of 2 in the terms corresponding to the ghosts because, even when the propagator in the Landau gauge corresponds to 3 propagating polarizations, the fourth polarization still contributes to the logarithm of the fluctuation determinant (after appropriate subtraction of an infinite piece going as $\log\xi$),  and the corresponding term is cancelled by half of the ghost contribution. The remaining ghost part and the three remaining gauge polarizations contribute as in \eqref{eq:VT}.\footnote{For example, given an Abelian field coupled to a complex scalar one would sum over 3 gauge polarizations, 2 real scalars, and subtract one ghost contribution; then effectively one has four degrees of freedom, as expected in either a broken phase (three massive gauge polarizations and a real scalar) or in the unbroken phase (2 massless gauge polarizations plus 2 real scalars).} The functions $J_B$ and $J_F$ are given by
 \begin{align}\label{eq:Js} \begin{aligned}
  J_B(x)=\int_0^\infty dy\, y^2\log\left[1-\exp(-\sqrt{x+y^2})\right],\\
   J_F(x)=\int_0^\infty dy\, y^2\log\left[1+\exp(-\sqrt{x+y^2})\right].
 \end{aligned}\end{align}
 In the ultrarelativistic limit, using the results $J_B(0)=-(8/7)J_F(0)=-\pi^4/45$ in Eq.~\eqref{eq:VT}, one recovers the standard pressure for a relativistic gas with the degrees of freedom in SMASH,
 \begin{align}
  -V_T\rightarrow p_{\rm rel}=\frac{\pi^2}{90}\,g_{\ast\rho,\rm rel} T^4, \quad g_{\ast\rho,\rm rel}=124.5.
 \end{align}
 
The naive value of $g_{\ast\rho}$ above will be modified  even in the high-temperature limit  as a consequence of the thermal corrections to the masses and the higher-order QCD corrections.
 Regarding the daisy resummed masses, the improved daisy resummation has to be applied to bosonic degrees of freedom that couple to fields whose masses may become larger than the temperature. For fields that remain light, the usual daisy resummation suffices.  The PQ phase transition gives masses to the real and imaginary components of $\sigma$, the vector quark $Q$, and the right-handed neutrinos. The former couple to the Higgs, the $\sigma$ fields, and the $SU(3)$ and hypercharge gauge bosons, for which we should in principle use the improved daisy resummation.  We will do so for the Higgs, $\sigma$, and hypercharged fields, while for the $SU(3)$ gauge bosons we will use an alternative treatment, to be discussed below, in order to allow for the inclusion of higher-order QCD corrections without incurring into double counting.
 
For a generic real scalar field $\phi_j$, the contributions of bosons and fermions to its thermal mass beyond the high-temperature expansion for scalar backgrounds $\phi_j=\bar\phi_j$ go as (see appendix~\ref{app:Daisy} for details)
 \begin{align}
 \Delta m^2_{\phi_j}(\bar\phi_i,T)\supset \sum_{B}\,\frac{T^2}{\pi^2}J_B'\left(\frac{m^2_{B}(\bar\phi_i)}{T^2}\right)\left.\frac{\partial m^2_{B}(\phi_i)}{\partial\phi_j^2}\right|_{\phi_i\rightarrow\bar\phi_i}-2\sum_{F}\,\frac{T^2}{\pi^2}\,J_F'\left(\frac{m^2_{F}(\bar\phi_i)}{T^2}\right)\left.\frac{\partial m^2_{F}(\phi_i)}{\partial \phi_j^2}\right|_{\phi_i\rightarrow\bar\phi_i}.
\end{align}
Again, the sum in $B$ goes over real scalars, and the sum in $F$ over Weyl fermions. The masses $ m^2_{B}(\phi_i)$ and $m^2_{F}(\phi_i)$ refer to an arbitrary background for all the scalars at zero temperature, while $ m^2_{B}(\bar\phi_i)$ and $m^2_{F}(\bar\phi_i)$ refer to the specific background of interest. In this way the
factors $\partial m^2_{B}(\phi_i)/\partial {\phi_j}^2$ and  $\partial m^2_{F}(\phi_i)/\partial{\phi_j}^2$ are just a convenient way to write the portal couplings of $\phi_j$ to other scalars, or the Yukawa couplings with fermions. Similarly, for an Abelian $U(1)$ field $G$ with gauge coupling $\tilde g$, one has the following contribution to the effective mass due to heavy Weyl fermions $F$ with charges $q_F$:
\begin{align}\begin{aligned}
  \Delta m^2_G(\bar\phi_i,T)\supset \sum_F\frac{\tilde g^2q_F^2T^2}{\pi^2} K_F\left(\frac{m^2_F(\bar\phi_i)}{T^2}\right),
\end{aligned}\end{align}
where
\begin{align}\label{eq:KF}
 K_F(x)=\int_0^\infty dy\, \frac{y^2 e^{\sqrt{x+y^2}}}{( e^{\sqrt{x+y^2}}+1)^2}.
\end{align}
For the contributions of fields that remain light in the range of temperatures considered (e.g. the SM fields), one can use the standard daisy resummation results.

As mentioned before one could also implement an improved daisy resummation for the gluons, but this would complicate the inclusion of additional QCD loop corrections without double counting. We have opted for including known three-loop QCD corrections in the contribution $V^{\rm QCD}$, and implementing decoupling of the heavy quarks $Q$ by interpolating between a seven flavour regime for $T>m_Q$ (with $m_Q$ denoting the mass of the heavy quarks) and a six flavour regime for $T<m_Q$. We use the results for $V^{\rm QCD}(T)=-p^{\rm QCD}(T)$ up to order $g^6_3$ of Ref.~\cite{Kajantie:2002wa}, which account for a variable number of massless flavours.
The interpolation is performed by considering a weighted sum of the seven and six flavour results. The weight of the seven flavour contribution  is taken as $J_F(m^2_Q/T^2)/J_F(0)$---which is zero at low temperature, and approaches one at high $T$---while the weight of the six flavour contribution is chosen as $1-J_F(m^2_Q/T^2)/J_F(0)$. A consistent inclusion of the three-loop QCD corrections requires knowing the QCD contributions to the beta function of $g_3$ in SMASH up to three loops. For this we use the results of Ref.~\cite{Pickering:2001aq}, which give the following three-loop contribution to the beta function $\beta_{g_3}=\mu\,\partial g_3/\partial\mu$ of SMASH:
\begin{align}
\Delta\beta_{g_3}= \frac{12629}{221184}\frac{g_3^7}{\pi^6}.
\end{align}

With the full thermal potential $V_{\rm eff}(\sigma,T)$ of Eq.~\eqref{eq:VT} computed as explained above, one can then obtain the quantities $g_{\ast\rho}$ and $g_{\ast s}$ from Eq.~\eqref{eq:gs_computation}. However, a remaining subtlety is that the formulae assume that all the relevant particles involved  share the same temperature. This neglects effects related to the loss of chemical equilibrium. For massive particles that decouple from the thermal bath, even though their effective temperature is expected to start to differ from that of the rest of the plasma, the effect is not relevant for the computation of $g_{\ast\rho}$ and $g_{\ast s}$, because decoupling ensures that the former quantities are only sensitive to the light species. However, this is not so for the axion degree of freedom emerging after the breaking of the PQ symmetry. As corresponds to a pseudo-Goldstone, the axion interacts very weakly, and at some point will fall out of chemical equilibrium with the rest of the plasma. After that, the axion stops interacting with the thermal bath, and the entropies of the former and the plasma are separately conserved, implying a deviation of the temperature of the axion bath---which will be referred to as $T_{\rm axion}$---from the temperature $T$ of the rest of the plasma.

In order to estimate the corrections arising from the axion decoupling, one needs to know the decoupling temperature $T_{\rm dec}$. In principle this can be determined by computing thermally averaged rates or solving appropriate Boltzmann equations,\footnote{Note that the setup considered here is somewhat different from 
what usually assumed in the studies of the thermal axion production such as Refs.~\cite{Masso:2002np,Graf:2010tv,Salvio:2013iaa}.
Based on the results of these analyses one may naively expect that axions decouple from the thermal plasma when the temperature drops below the following critical value~\cite{Graf:2010tv},
\begin{align}
T_{\rm dec} \sim 1.7\times 10^9\,\mathrm{GeV}\,\left(\frac{f_a}{10^{11}\,\mathrm{GeV}}\right)^{2.246}.
\label{eq:T_dec_usual}
\end{align}
However, in SMASH we cannot apply this estimate since the critical temperature of the PQ phase transition is comparable or even lower than the value of $T_{\rm dec}$ shown above.
In other words, axions have not yet emerged as a Goldstone degree of freedom at the would-be decoupling temperature~\eqref{eq:T_dec_usual}, and decouple only when 
the PQ symmetry breaking field $\sigma$ acquires a sufficiently large expectation value.
This fact motivates us to consider $T_{\rm dec}$ to be lower than the estimate given by Eq~\eqref{eq:T_dec_usual}, as discussed in the text above Eq.~\eqref{eq:delta_def}.} 
but here we adopt an approximation consisting in estimating the temperature at which the phase transition can be considered complete. While the usual critical temperature $T_c$ corresponds to the point at which a new vacuum starts to arise, for a second-order phase transition the vacuum at $T_c$ still lies at the origin, and the scale of symmetry breaking only emerges at lower temperatures. An indicator of the emergence of the PQ breaking scale is the trace of the stress-energy momentum, which is zero in a plasma of  massless particles but acquires peaks when new dimensionful scales emerge and the associated massive degrees of freedom are not yet thermally decoupled. Due to this, we will estimate $T_{\rm dec}$ as the temperature 
at which the trace of the stress-energy tensor exhibits a local maximum. For this we will consider the following dimensionless quantity (called the trace anomaly),
\begin{align}
 \Delta(T) = \frac{T^\mu_\mu}{T^4}=\frac{\rho-3p}{T^4}=\frac{1}{T^4}\left(4V_{\rm eff,min}(T)-T\frac{\partial V_{\rm eff,min}(T)}{\partial T}\right).
 \label{eq:delta_def}
\end{align}

Once $T_{\rm dec}$ is determined from the maximum of $\Delta$, the values of 
 $g_{\ast\rho}$ and $g_{\ast s}$ accounting for the axion decoupling can be calculated as follows. First, we may denote the quantities  obtained from Eq.~\eqref{eq:gs_computation} under the chemical equilibrium assumption as $g^{\rm eq}_{\ast\rho}$ and $g^{\rm eq}_{\ast s}$. As a massless axion contributes one unit to both $g^{\rm eq}_{\ast\rho}$ and $g^{\rm eq}_{\ast s}$, we can obtain the entropy and energy density of the axion (with temperature $T_{\rm axion}$) and of the thermal bath with the axion excluded (with temperature $T$) as:
\begin{align}\label{eq:rhosbathaxion}\begin{aligned}
 \rho^{\rm axion}=&\,\frac{\pi^2}{30}\,T^4_{\rm axion}, &\rho^{\rm bath}=&\,\frac{\pi^2}{30}\,g_{\ast\rho}^{\rm bath}\,T^4, &g^{\rm bath}_{\ast\rho}=&\,g^{\rm eq}_{\ast\rho}-1,\\
s^{\rm axion}=&\,\frac{2\pi^2}{45}\,T^3_{\rm axion}, & s^{\rm bath}=&\,\frac{2\pi^2}{45}\,g_{\ast s}^{\rm bath}\,T^3, &g^{\rm bath}_{\ast s}=&\,g^{\rm eq}_{\ast s}-1.
\end{aligned}\end{align}
Below the decoupling temperature $T_{\rm dec}$, the separate conservation of $s^{\rm axion}$ and $s^{\rm bath}$ implies
\begin{align}\label{eq:Taxion}
 T_{\rm axion}=\left\{\begin{array}{cl}
                        T,&T\geq T_{\rm dec},\\
                       \left(\frac{g^{\rm bath}_{\ast s}(T)}{g^{\rm bath}_{\ast s}(T_{\rm dec})}\right)^{\frac{1}{3}} T, &T< T_{\rm dec}.
                      \end{array}\right.
\end{align}
Then the final values of $g_{\ast\rho}$ and $g_{\ast s}$ follow from the total energy density and entropy, $\rho=\rho^{\rm bath}+\rho^{\rm axion}$, $s=s^{\rm bath}+s^{\rm axion}$, through Eqs.~\eqref{eq:rhosT} and \eqref{eq:rhosbathaxion}:
\begin{align}
 g_{\ast\rho}=g^{\rm eq}_{\ast\rho}-1+\left(\frac{T_{\rm axion}}{T}\right)^{4},\qquad g_{\ast s}= g^{\rm eq}_{\ast s}-1+\left(\frac{T_{\rm axion}}{T}\right)^3,
\label{eq:gsrhosbathaxion}
\end{align}
where $T_{\rm axion}/T$ is to be computed using \eqref{eq:Taxion}. 

In Fig.~\ref{fig:eoszoom}, we plot the estimate of $g_{\ast\rho}$ and $g_{\ast s}$ obtained based on the procedure described above.
We also show the uncertainty of $g_{\ast\rho}$ and $g_{\ast s}$ estimated by varying the renormalization scale in the range $\mu \in (0.5\dots 2)m_{\rho}$, 
where $m_{\rho} = \sqrt{2\lambda_{\sigma}}v_{\sigma}$ is the mass of the radial direction $\rho$ of the PQ field, 
and the value of the unknown coefficient $q_c$ in the order $g_3^6$ result for the QCD pressure~\cite{Kajantie:2002wa} in the range $q_c \in (-5000\dots + 5000)$.
The uncertainty from the $\mathcal{O}(g_3^6)$ QCD pressure becomes less important at higher temperatures 
since the QCD coupling becomes  weaker and such a contribution is more suppressed at higher energies.
As shown in Fig.~\ref{fig:eoszoom}, the values of $g_{\ast\rho}$ and $g_{\ast s}$ start to fall off when the temperature becomes lower than 
the critical temperature $T_c$ of the PQ phase transition.\footnote{The critical temperature of the PQ phase transition shown in 
Figs.~\ref{fig:eoszoom},~\ref{fig:SMmatch}
and Table~\ref{tab:T_c_and_T_dec} is defined as the temperature at which the expectation value of the $\sigma$ field starts to deviate from
$\langle|\sigma|^2\rangle = 0$, which is found numerically by minimizing the full effective potential~\eqref{eq:V_eff_full} 
rather than using the approximate formula~\eqref{eq:T_c_analytical}.}
Subsequently, there are step-like changes in $g_{\ast\rho}$ and $g_{\ast s}$ when the temperature becomes 
about an order of magnitude lower than $T_c$, which corresponds to the decoupling of heavy fermions $Q$ and $N_i$.
In the next section, we will see that such a change in $g_{\ast\rho}$ and $g_{\ast s}$ leads to an observable signature in the spectrum of primordial GWs.

\begin{figure}[h]
\centering
\includegraphics[width=0.5\linewidth]{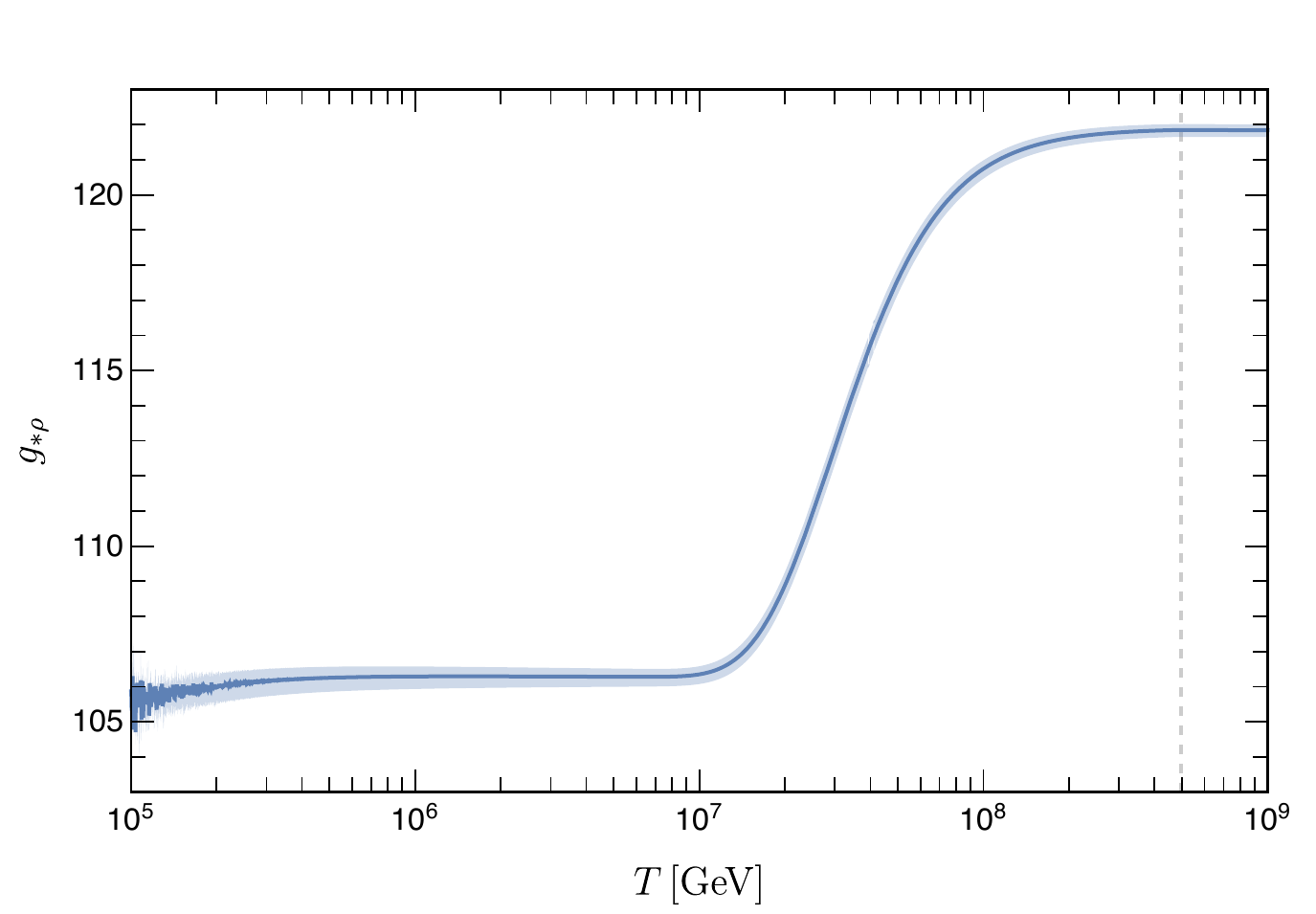}\includegraphics[width=0.5\linewidth]{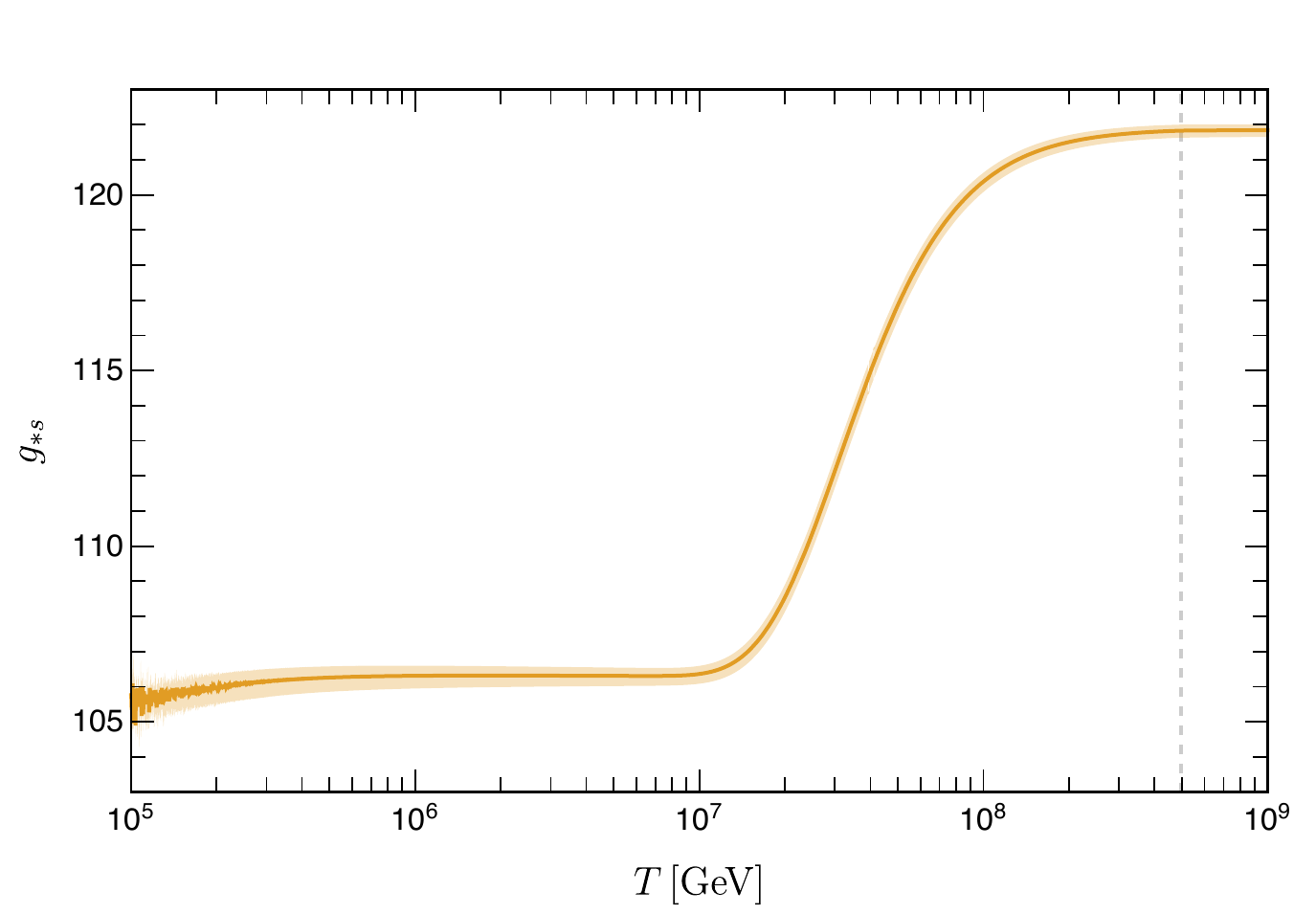}\\\includegraphics[width=0.5\linewidth]{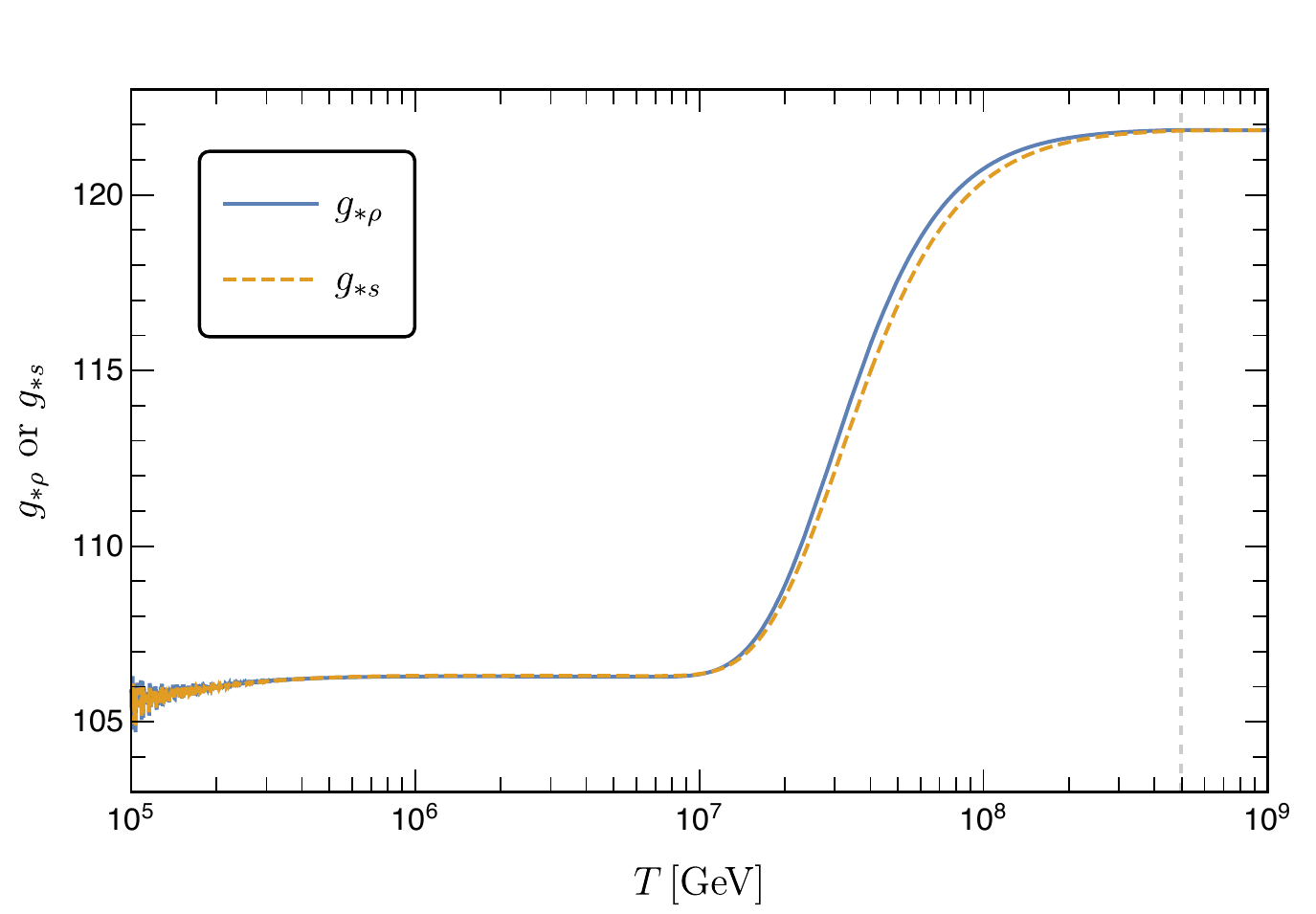}
\caption{
Temperature dependence of $g_{\ast\rho}(T)$ (top left) and $g_{\ast s}(T)$ (top right) obtained from Eq.~\eqref{eq:gsrhosbathaxion}.
Shaded regions represent the uncertainty due to the choice of the renormalization scale and estimate of the QCD corrections (see text).
Note that these regions do not necessary indicate the largest source of uncertainty, since we expect that there might be further corrections if we go beyond the sudden decoupling approximation to estimate the contribution of relativistic axions.
Gray dashed lines correspond to the critical temperature of the PQ phase transition. In the bottom panel, central values of $g_{\ast\rho}$ and $g_{\ast s}$ are compared directly. 
In these figures, parameter values are taken according to benchmark point 1 in Table~\ref{tab:BMP_parameters}.}
\label{fig:eoszoom}
\end{figure}

Practically, the calculation based on the thermal potential~\eqref{eq:V_eff_full} exhibits numerical instabilities at low temperatures, since it involves cancellation of large numbers. 
This is clearly seen in Fig.~\ref{fig:eoszoom}, where the values of $g_{\ast\rho}$ and $g_{\ast s}$ fluctuate a lot at $T=\mathcal{O}(10^5)\,\mathrm{GeV}$.
This fact prevents us from tracing the decoupling of the radial component of the PQ field with mass $m_{\rho} \sim 10^5$--$10^7\,\mathrm{GeV}$ 
and evaluating the effective relativistic degrees of freedom at lower temperatures. 
To avoid this issue, we evaluate thermodynamic quantities at lower temperatures by adding the contribution of the free $\rho$ particle, 
which can be estimated analytically, to $\Delta$ for the SM:
\begin{align}
\Delta(T)&= \Delta_{\rm SM}(T) + \Delta_{\rho}(T),\label{eq:Delta_SM_rho}\\
\Delta_{\rho}(T) &= \left.\frac{x}{\pi^2}J_B'(x)\right|_{x\,=\,m_{\rho}^2/T^2},
\end{align}
where $\Delta_{\rm SM}$ denotes the contribution of the SM particles.
Here we use the result of Ref.~\cite{Saikawa:2018rcs} to evaluate $\Delta_{\rm SM}$ and match the SMASH result~\eqref{eq:delta_def} to Eq.~\eqref{eq:Delta_SM_rho}.
As shown in Fig.~\ref{fig:SMmatch}, the two results indeed agree with each other, which allows us to interpolate between the SM and SMASH results by using Eq.~\eqref{eq:Delta_SM_rho}.
The interpolation is performed by applying a Gaussian filter to smooth out the fluctuations in the SMASH result~\eqref{eq:delta_def} at low temperatures and 
switching from Eq.~\eqref{eq:delta_def} to Eq.~\eqref{eq:Delta_SM_rho} at a point where the central values of two results cross each other.
After obtaining the interpolated function of $\Delta(T)$, we estimate the pressure $p(T)$ by integrating $\Delta(T)$ in a similar way to Ref.~\cite{Saikawa:2018rcs}, 
to obtain $g_{\ast\rho}$ and $g_{\ast s}$ at arbitrary temperature.

\begin{figure}[h]
\centering
\includegraphics[width=0.5\linewidth]{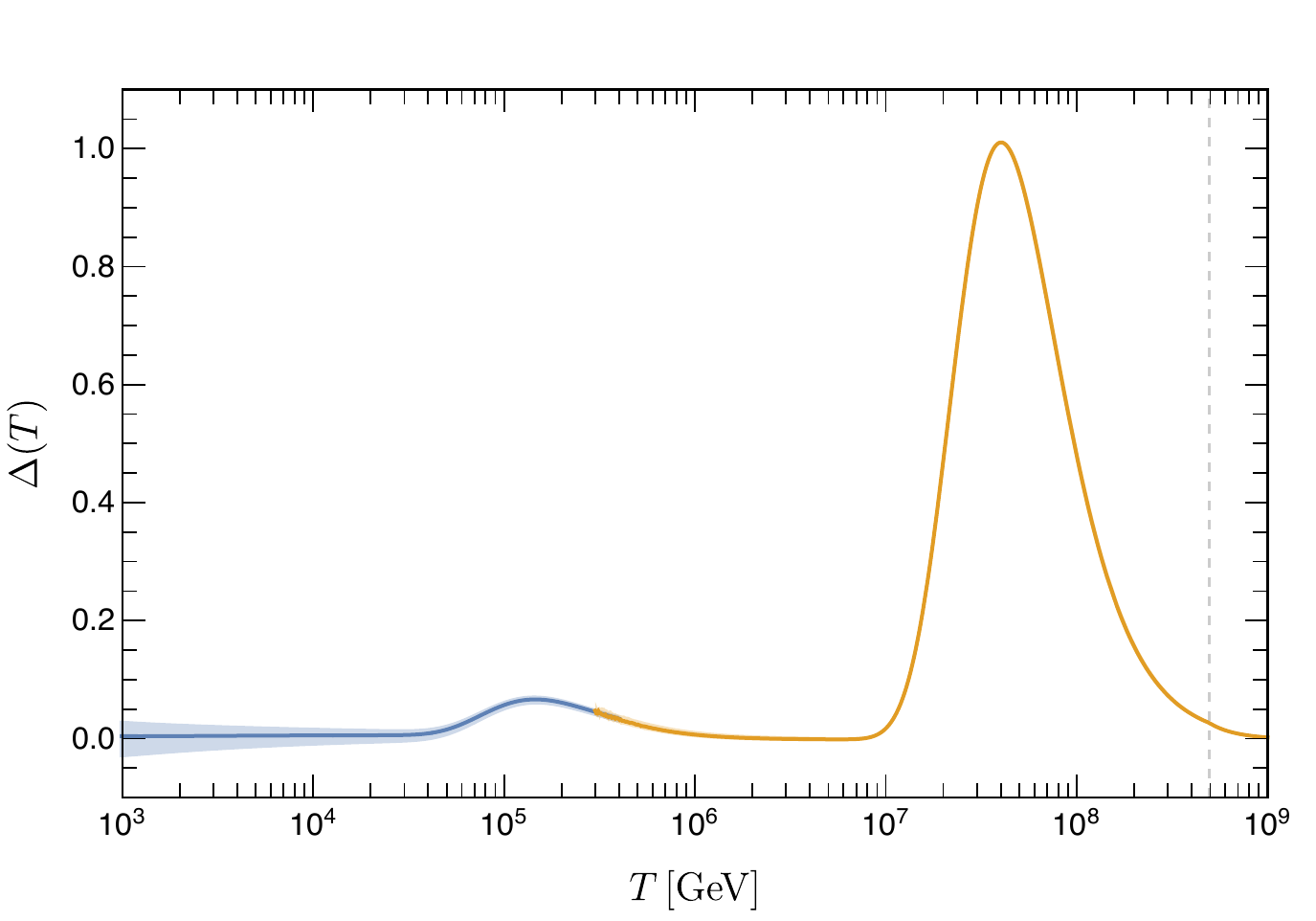}\includegraphics[width=0.5\linewidth]{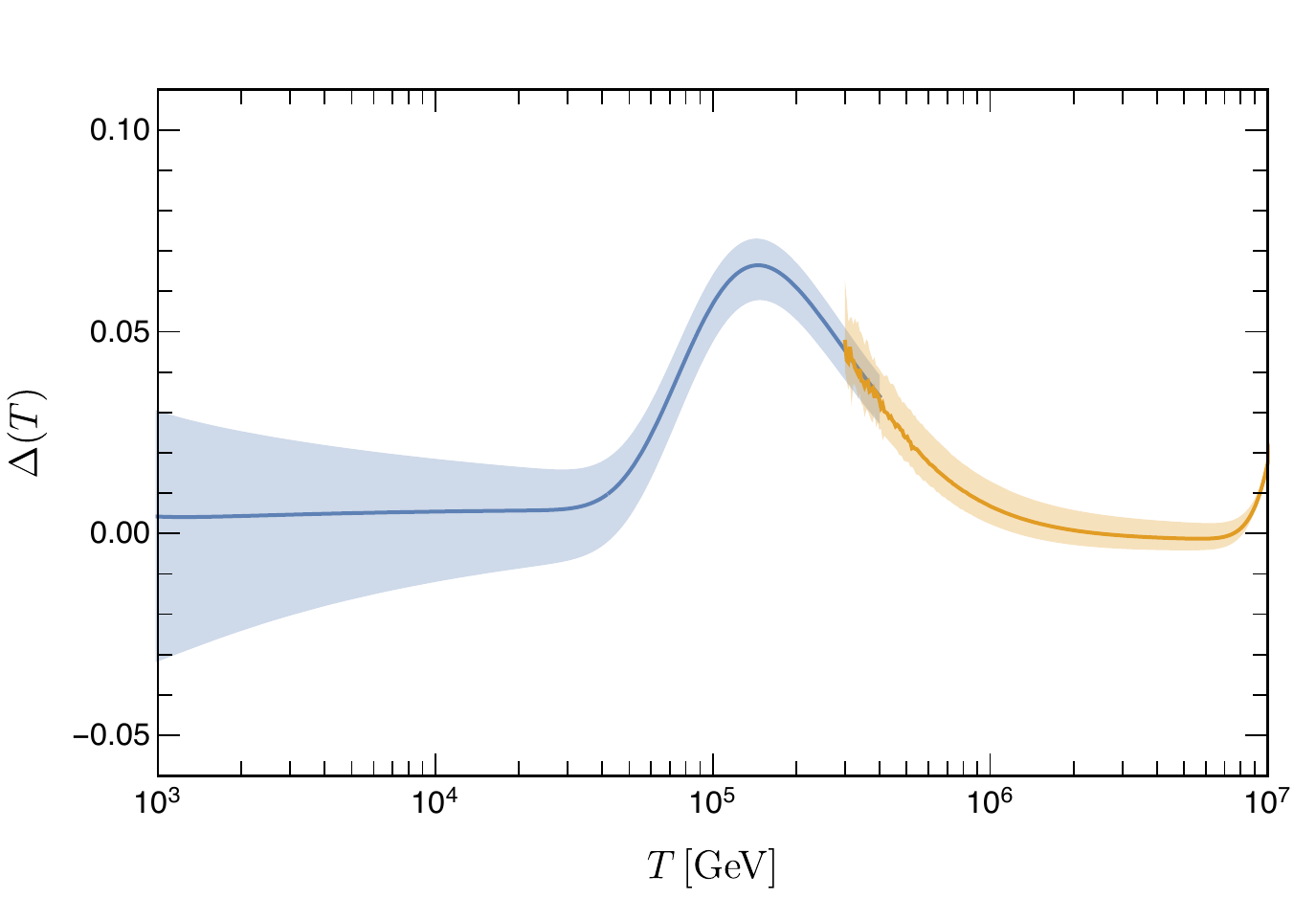}
\caption{
Temperature dependence of $\Delta(T)$ evaluated according to Eq.~\eqref{eq:Delta_SM_rho} (blue) and Eq.~\eqref{eq:delta_def} (orange).
Right panel just shows a zoomed plot of the left panel. Shaded regions represent the uncertainty due to the choice of the renormalization scale and estimate of the QCD corrections (see text),
and the gray dashed line corresponds to the critical temperature of the PQ phase transition.
In these figures, parameter values are taken according to benchmark point 1 in Table~\ref{tab:BMP_parameters}.}
\label{fig:SMmatch}
\end{figure}

To finalize, we provide four SMASH benchmark points associated with different values of the effective inflationary quartic coupling at high scales. These imply different effective quartic couplings $\tilde\lambda_\sigma$ along the inflationary valley  (see Eq.~\eqref{genpotential}). Under the assumption of a given value of the non-minimal gravitational coupling $\xi_\sigma$, and under the requirement of a consistent reheating history, the value of the tensor-to-scalar ratio $r(k_*)$ during inflation becomes fixed in terms of $\tilde\lambda_\sigma$ \cite{Ballesteros:2016euj,Ballesteros:2016xej}. The chosen benchmark points are summarized in Table~\ref{tab:BMP_parameters}. In all four cases, we take the scale of the PQ symmetry breaking to be $v_{\sigma} = 1.2 \times 10^{11}\,\mathrm{GeV}$. In Table~\ref{tab:T_c_and_T_dec}, we also show the values of the critical temperature of the PQ phase transition $T_c$ and the axion decoupling temperature $T_{\rm dec}$, which corresponds to the local maximum of $\Delta$, for each benchmark point.

\begin{table}
\begin{center}
\begin{tabular}{c|c|c|c|c}
{\rm Benchmark point} &  1 & 2 & 3 & 4\\
\hline
\hline\rule{0pt}{17pt}
$r(0.002 \text{ Mpc}^{-1})$    & 0.048                &  0.0096              &  0.0068           & 0.0037\\[5pt]
\hline \rule{0pt}{17pt}
$n_s(0.002 \text{ Mpc}^{-1})$    & 0.9642              &  0.9663            &  0.9665           & 0.9666\\[5pt]
\hline \rule{0pt}{17pt}
$\phi_*/M_P$                 &     22                 &     18                &     16              & 8.4\\[5pt]
\hline\rule{0pt}{17pt}
$\xi_\sigma(\phi_*)$           & 0.0096               &  0.079               & 0.14              & 1.0\\[5pt]
\hline\rule{0pt}{17pt}
$\tilde\lambda_\sigma(\phi_*)$ & $9.1\times10^{-13}$  & $9.0\times10^{-12}$  &$2.0\times10^{-11}$ & $5.3\times10^{-10}$ \\[5pt]
\hline\rule{0pt}{17pt}
$\lambda_\sigma(M_P)$          &  $4.4\times10^{-12}$ & $1.4\times10^{-10}$  &$5.0\times10^{-11}$  & $4.4\times10^{-9}$\\[5pt]
\hline\rule{0pt}{17pt}
$\lambda_{H\sigma}(M_P)$       & $-1.5\times10^{-6}$  & $-6.0\times10^{-6}$  & $-6.5\times10^{-6}$& $-2.9\times10^{-5}$\\[5pt]
\hline  \rule{0pt}{17pt}
$\lambda_H(M_P)$               &  0.63                & 0.26                 &1.2                  & 0.21\\[5pt]
\hline\rule{0pt}{17pt}
$y(M_P)$                       & 0.00056              & 0.0014               & 0.00086             & 0.0027\\[5pt]
\hline\rule{0pt}{17pt}
$Y_{ii}(M_P)$                  & 0.0011               & 0.0025               & 0.0016              & 0.0045\\[5pt]
\hline
\end{tabular}\\
\end{center}
\caption{Parameter values for the chosen SMASH benchmark points. We take $v_{\sigma} = 1.2 \times 10^{11}\,\mathrm{GeV}$ in all four benchmark points.  $\phi_*$ denotes the value of the inflaton field (in the Jordan frame) when the Planck pivot scale of 0.002 ${\rm Mpc}^{-1}$ crosses the horizon.}
\label{tab:BMP_parameters}
\end{table}

\begin{table}
\begin{center}
\begin{tabular}{c||c|c}
{\rm Benchmark point} & $T_c\,[10^9\,\mathrm{GeV}]$ & $T_{\rm dec}\,[10^7\,\mathrm{GeV}]$\\
\hline
1 & 0.48--0.51 & 3.9--4.1\\
2 & 1.25--1.31 & 8.7--9.1\\
3 & 1.34--1.44 & 5.6--5.8\\
4 & 4.50--5.20 & 15.6--16.1
\end{tabular}
\end{center}
\caption{Values of the critical temperature of the PQ phase transition $T_c$
and axion decoupling temperature $T_{\rm dec}$ for the chosen SMASH benchmark points.
The range of values corresponds to the uncertainty due to the choice of the renormalization scale.}
\label{tab:T_c_and_T_dec}
\end{table}

In Fig.~\ref{fig:eosfull}, we show the estimate of $g_{\ast\rho}$, $T_{\rm axion}/T$, and $\Delta$ for the chosen SMASH benchmark points. 
We see that the value of the effective relativistic degrees of freedom in SMASH becomes larger than that in the SM at high temperatures.
In particular, there are two step-like features at $T\sim m_{\rho}$ (threshold of $\rho$ particle) and at $T\sim T_{\rm dec}$ (threshold of heavy fermions $Q$ and $N_i$).
These events correspond to the two peaks in $\Delta(T)$, which are shown in the bottom panel of Fig.~\ref{fig:eosfull}.
Note that the ordering of the benchmark points (and hence the ordering of the value of the tensor-to-scalar ratio $r$) does not imply
an analogous ordering of the locations of the peaks, since $r$ is related to the effective inflationary coupling $\tilde{\lambda}_{\sigma}$,
which itself depends on a combination of three coupling parameters (see Eq.~\eqref{genpotential}).
Furthermore, the overall values of $g_{\ast\rho}$ and $g_{\ast s}$ remain slightly larger than the SM values even at lower temperatures, 
because of the extra contribution from relativistic axions (see Eq.~\eqref{eq:gsrhosbathaxion}).
The contribution of the relativistic axions is proportional to a power of $T_{\rm axion}/T$, which decays at low temperature like $\propto [g_{\ast s}^{\rm bath}(T)]^{1/3}$
as shown in the top right panel of Fig.~\ref{fig:eosfull}.

\begin{figure}[h]
\centering
\includegraphics[width=0.5\linewidth]{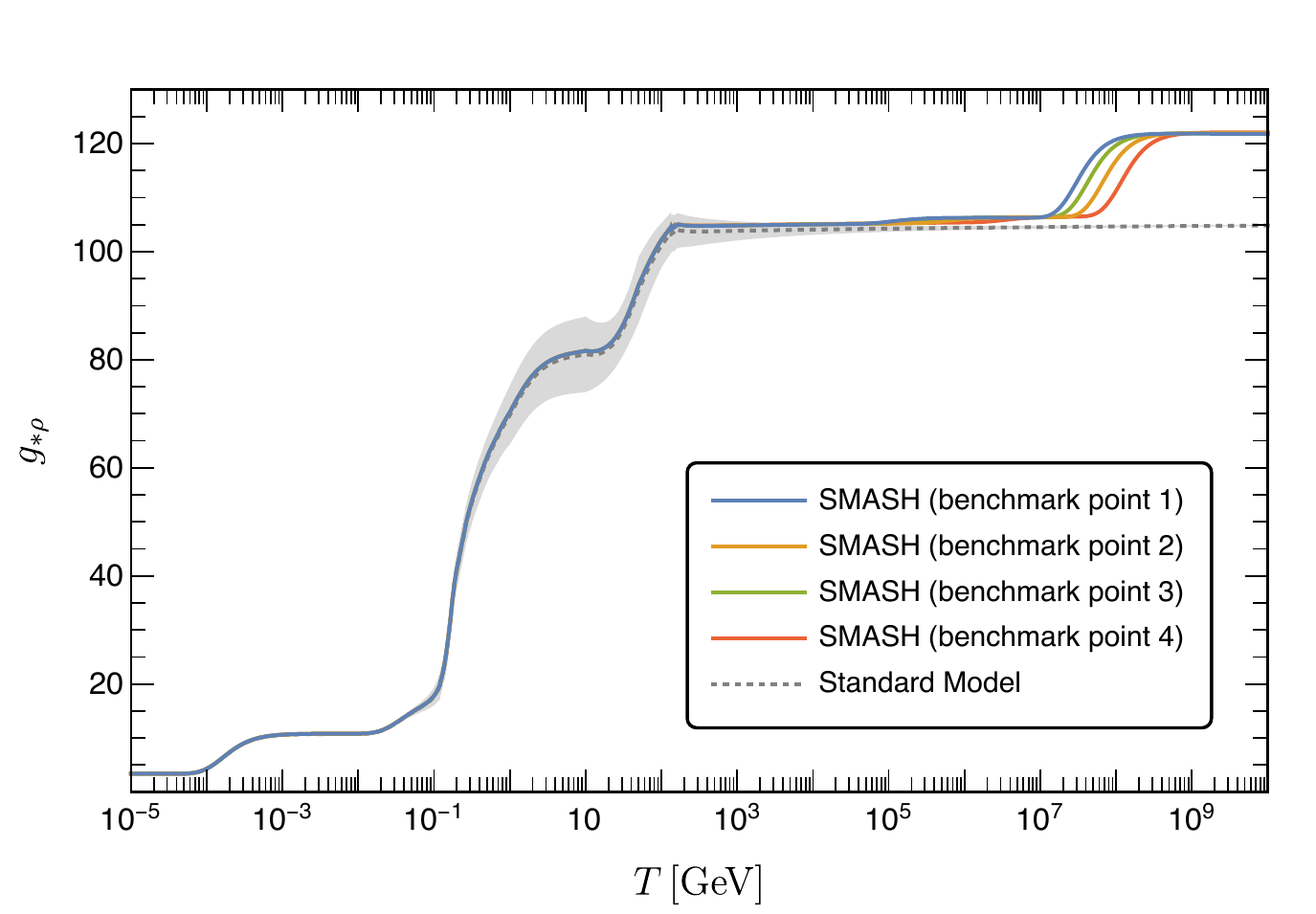}\includegraphics[width=0.5\linewidth]{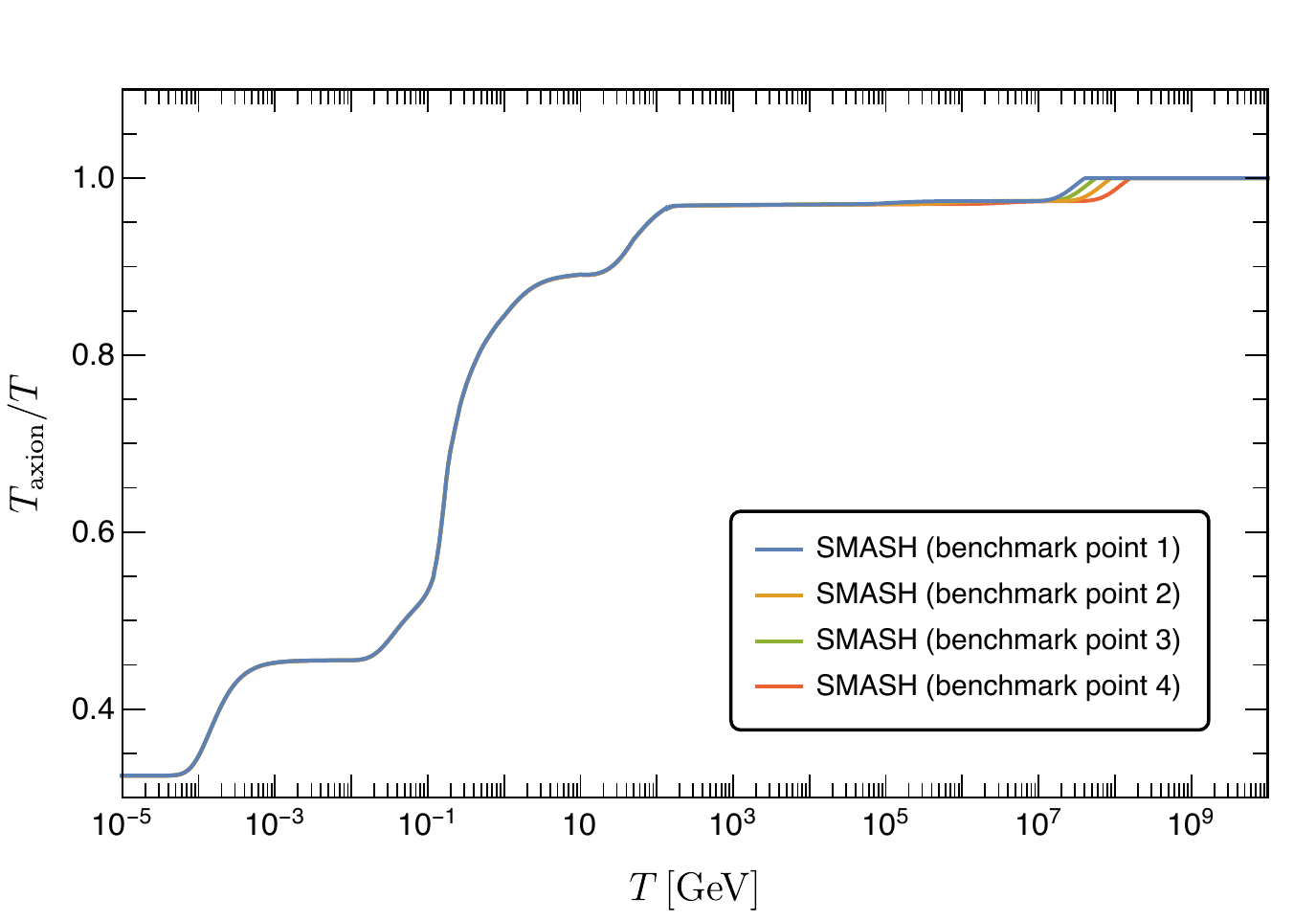}\\\includegraphics[width=0.5\linewidth]{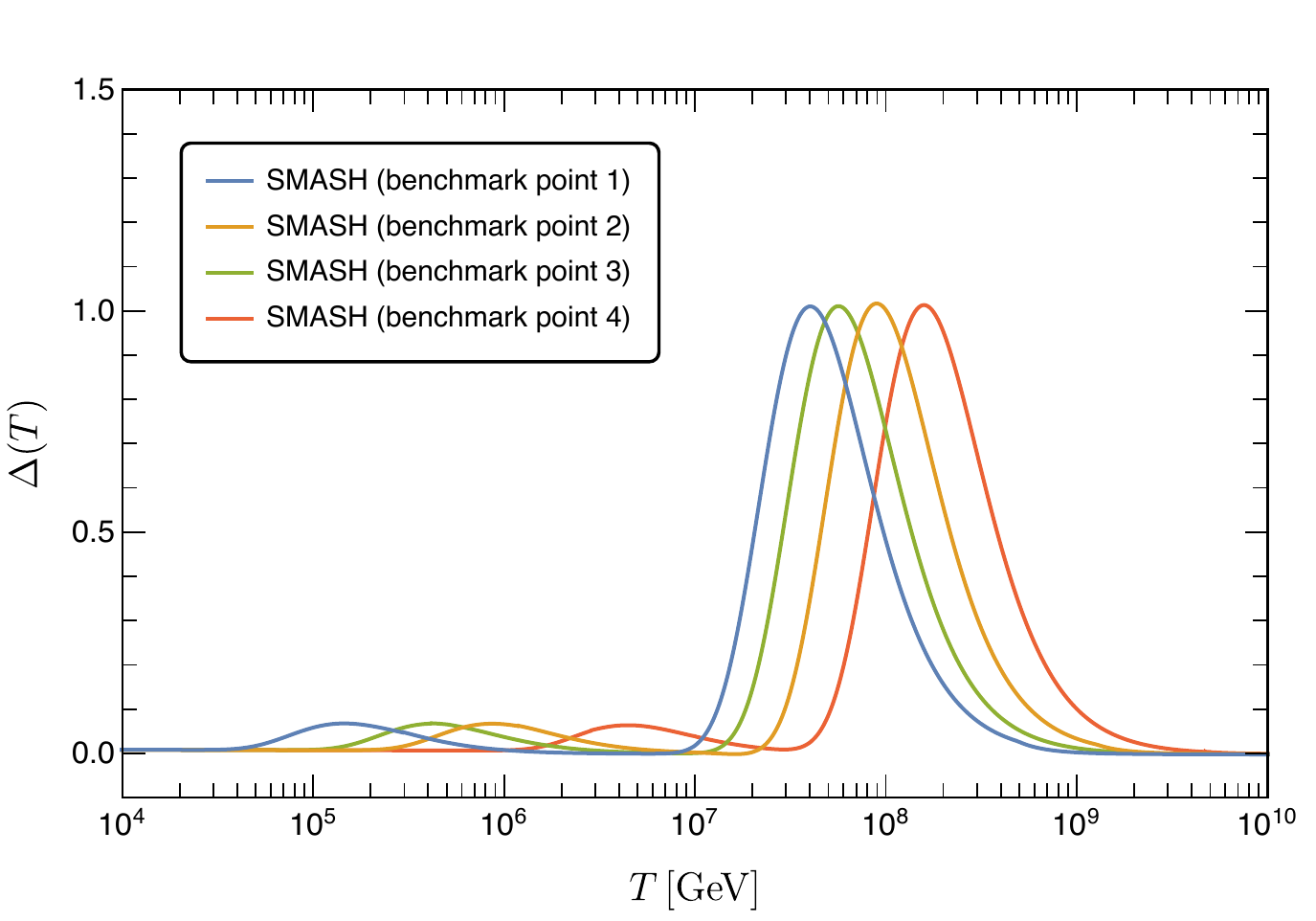}
\caption{Temperature dependence of $g_{\ast\rho}(T)$ (top left), $T_{\rm axion}/T$ (top right), and $\Delta(T)$ (bottom) for the chosen SMASH benchmark points.
For the plots of $g_{\ast\rho}(T)$, we also show the SM result (gray dashed line) and its uncertainty (gray shaded region) estimated in Ref.~\cite{Saikawa:2018rcs}.}
\label{fig:eosfull}
\end{figure}

Before closing this section, let us estimate the asymptotic values of $g_{\ast\rho}$ and $g_{\ast s}$ after the neutrino decoupling, which are relevant to the present observations.
The conservation of entropy implies that the ratio of the effective temperatures of neutrinos and axions to photon temperature after the neutrino decoupling are respectively
\begin{align}
\frac{T_{\nu}}{T_{\gamma}} = \left(\frac{4}{11}\right)^{\frac{1}{3}}, \quad \frac{T_{\rm axion}}{T_{\gamma}} = \left(\frac{43}{11g_{\ast s}^{\rm bath}(T_{\rm dec})}\right)^{\frac{1}{3}}.
\end{align}
By using these formulae, the asymptotic values of $g_{\ast\rho}$ and $g_{\ast s}$ after the neutrino decoupling can be estimated as
\begin{align}
g_{\ast\rho,0} &= 2 + \frac{21}{4}\left(\frac{4}{11}\right)^{\frac{4}{3}} + \left(\frac{43}{11g_{\ast s}^{\rm bath}(T_{\rm dec})}\right)^{\frac{4}{3}} \simeq 3.36 + 0.01\,\left(\frac{114}{g_{\ast s}^{\rm bath}(T_{\rm dec})}\right)^{\frac{4}{3}},
\label{eq:gsrho0_SMASH}\\
g_{\ast s,0} &= \frac{43}{11}\left(1+\frac{1}{g_{\ast s}^{\rm bath}(T_{\rm dec})}\right) \simeq 3.91 + 0.03\,\left(\frac{114}{g_{\ast s}^{\rm bath}(T_{\rm dec})}\right).
\label{eq:gss0_SMASH}
\end{align}
The former can be recast in terms of an excess in effective number of neutrinos,
\begin{align}
\Delta N_{\rm eff} = \frac{4}{7}\left(\frac{T_{\rm axion}}{T_{\nu}}\right)^4 \simeq 0.0245\,\left(\frac{114}{g_{\ast s}^{\rm bath}(T_{\rm dec})}\right)^{\frac{4}{3}}.
\label{eq:DeltaNeff_SMASH}
\end{align}
The excess in $N_{\rm eff}$ may be probed in future CMB and large scale structure observations~\cite{Baumann:2017gkg}.
We numerically find that the chosen SMASH benchmark points predict $g_{\ast s}^{\rm bath}(T_{\rm dec}) \simeq 114$,
and hence $g_{\ast\rho,0} \simeq 3.37$, $g_{\ast s,0}\simeq 3.94$, and $\Delta N_{\rm eff} \simeq 0.0245$.
These values are slightly larger than those predicted in the SM, $g_{\ast\rho,0} \simeq 3.36$, $g_{\ast s,0}\simeq 3.91$, and $\Delta N_{\rm eff} = 0$.\footnote{Strictly speaking, 
when we take account of the feeble interaction of neutrinos with the electromagnetic plasma and the leading order quantum electrodynamics corrections,
the asymptotic values of the effective relativistic degrees of freedom in the SM become $g_{\ast\rho,0} \simeq 3.38$ and $g_{\ast s,0}\simeq 3.93$~\cite{Saikawa:2018rcs}
(see also Refs.~\cite{Mangano:2001iu,Mangano:2005cc,deSalas:2016ztq} for a precision calculation of the neutrino decoupling
and Refs.~\cite{Bennett:2019ewm,Escudero:2020dfa,Akita:2020szl,Froustey:2020mcq} for more recent works), 
which are slightly larger than the commonly used values $g_{\ast\rho,0} \simeq 3.36$ and $g_{\ast s,0} \simeq 3.91$.
The SMASH predictions shown in Eqs.~\eqref{eq:gsrho0_SMASH}--\eqref{eq:DeltaNeff_SMASH} may also be subjected to similar corrections, 
whose precise estimation is beyond the scope of this paper.}
As the amplitude of GWs observed today depends on $g_{\ast s,0}$ (see Eq.~\eqref{Omega_gw_est}),
we use the value estimated based on Eq.~\eqref{eq:gss0_SMASH} when we evaluate the GW signatures in the next section.

\section{Spectrum of gravitational waves}
\label{sec:GW}
\setcounter{equation}{0}

Now we compute the spectrum of GWs predicted in SMASH by using the ingredients obtained in the previous sections. 
The GWs can be described by spatial metric perturbations $h_{ij}$ on the Friedmann-Robertson-Walker (FRW) background,
\begin{align}
ds^2 = -dt^2 + a^2(t)(\delta_{ij} + h_{ij})dx^idx^j,
\end{align}
where it is assumed that $h_{ij}$ satisfy the transverse-traceless conditions, $h^i_i = \partial^i h_{ij} = 0$.
We expand $h_{ij}$ in terms of their Fourier modes,
\begin{align}
h_{ij}(t,{\bf x}) = \sum_{\lambda}\int\frac{d^3k}{(2\pi)^3}h^{\lambda}(t,{\bf k})\epsilon^{\lambda}_{ij}({\bf k})e^{i{\bf k \cdot x}},
\label{eq:h_ij_def}
\end{align}
where $\lambda = +,\times$ represents two independent polarization states, and $\epsilon^{\lambda}_{ij}({\bf k})$ are the spin-2
polarization tensors that satisfy $\sum_{ij}\epsilon^{\lambda}_{ij}(\epsilon^{\lambda'}_{ij})^* = 2\delta^{\lambda\lambda'}$.
In the analysis of the spectrum of the primordial GWs, it will be convenient to introduce a dimensionless quantity $X(t,k)$ defined as
\begin{align}
h^{\lambda}(t,{\bf k}) = h^{\lambda}_{{\bf k},p} X(t,k),
\end{align}
where $h^{\lambda}_{{\bf k},p}$ represents the amplitudes of primordial tensor perturbations that are fixed when the corresponding modes cross outside the horizon,
and $X(t,k)$ satisfies $X(t,k) \to 1$ for $k\ll aH$.
The former is related to the primordial tensor power spectrum in Eq.~\eqref{eq:tensor_power_spectrum_def} as
\begin{align}
\langle h_{p,ij}(t,{\bf x}) h_{p,ij}(t,{\bf x}) \rangle = \int\frac{dk}{k}\Delta_t^2(k),
\end{align}
where $h_{p,ij}(t,{\bf x})$ are the primordial tensor fields given by Eq.~\eqref{eq:h_ij_def} with $h^{\lambda}(t,{\bf k})$ replaced by $h^{\lambda}_{{\bf k},p}$,
and $\langle\dots\rangle$ denotes the ensemble average.
It can also be written as
\begin{align}
\Delta_t^2(k) = \frac{2k^3|h_p(k)|^2}{\pi^2},
\end{align}
where $|h_p(k)|^2$ is defined by 
\begin{align}
\langle h^{\lambda*}_{{\bf k},p}h^{\lambda'}_{{\bf k'},p} \rangle = (2\pi)^3\delta^{\lambda\lambda'}\delta^{(3)}({\bf k}- {\bf k'})|h_p(k)|^2.
\end{align}
On the other hand, the quantity $X(t,k)$ describes the evolution of GWs after inflation, and it can be well-approximated by the WKB solution 
$X \propto a^{-1}e^{\pm ik\tau}$ for $k \gg aH$.

Substituting Eq.~\eqref{eq:h_ij_def} into the formula for the energy density of GWs~\cite{Maggiore:1900zz},
\begin{align}
\rho_{\rm gw}(t) = \frac{1}{32\pi G}\langle\dot{h}_{ij}(t,{\bf x})\dot{h}_{ij}(t,{\bf x})\rangle,
\end{align}
where $G$ is the Newton's gravitational constant,
we see that the spectrum of GWs $\Omega_{\rm gw}(f)$ can be represented as the product of the primordial tensor power spectrum $\Delta_t^2(f)$ 
and the transfer function $\mathcal{T}_0(f)$ as shown in Eq.~\eqref{Omega_gw_def}. 
Here, the transfer function is calculated from a derivative of the dimensionless quantity $X(\tau,k)$ with respect to conformal time $d\tau = dt/a$,
\begin{align}
\mathcal{T}_0(f) = \frac{1}{12a_0^2H_0^2}\left[\frac{dX(\tau_0,k)}{d\tau}\right]^2,
\label{eq:transfer_function}
\end{align}
where the comoving wavenumber $k$ is related to the frequency via $f=k/(2\pi a_0)$.

At linear order in perturbation theory, the evolution of $X(\tau,k)$ is described by the following integro-differential equation~\cite{Weinberg:2003ur,Watanabe:2006qe},
\begin{align}
\frac{d^2X(u)}{du^2} + \frac{2}{a(u)}\frac{da(u)}{du}\frac{dX(u)}{du} + X(u)
= -24\sum_{i=\gamma,\nu,a}F_i(u)\left[\frac{1}{a(u)}\frac{da(u)}{du}\right]^2\int^u_{u_i}dU\left[\frac{j_2(u-U)}{(u-U)^2}\right]\frac{dX(U)}{dU},
\label{eq:int_diff_eq_for_chi}
\end{align}
where $u = k\tau$ and $j_n(z)$ is the spherical Bessel function of the first kind.
The right-hand side of Eq.~\eqref{eq:int_diff_eq_for_chi} represents the collisionless damping effect
due to free-streaming particles~\cite{Weinberg:2003ur}, which is proportional to the fraction of the energy density in these particles,
\begin{align}
F_i(u) \equiv \frac{\rho_i(u)}{\rho_{\rm crit}(u)}.
\label{eq:F_i_def}
\end{align}
Here we consider photons ($i=\gamma$), neutrinos ($i=\nu$), and relativistic axions ($i=a$) as the sources of the damping effect. 
The lower limit of the integration $u_i$ is taken as the time at which the corresponding particles decouple from the thermal bath. Following Ref.~\cite{Saikawa:2018rcs}, we take the times corresponding to $T=3000\,K$ and $T=2\,\mathrm{MeV}$ for photons and neutrinos, respectively.
For axions, we use $T=T_{\rm dec}$, which is determined from the local maximum of $\Delta$ defined in Eq.~\eqref{eq:delta_def}.
It turned out that the contribution of free-streaming neutrinos leads to the suppression of the amplitude of GWs by $\lesssim 35\,\%$ at frequencies $10^{-16}\,\mathrm{Hz} \lesssim f \lesssim 10^{-10}\,\mathrm{Hz}$, and that of photons leads to the additional damping by $\lesssim 14\,\%$ at $f \sim 10^{-17}\,\mathrm{Hz}$~\cite{Saikawa:2018rcs}. In SMASH, there is an extra contribution from relativistic axions, which we study in this section.\footnote{The effect of relativistic axions on the spectrum of primordial GWs in a general context was discussed in Refs.~\cite{Jinno:2012xb,Jinno:2013xqa}.}

We numerically solve Eq.~\eqref{eq:int_diff_eq_for_chi} with initial conditions $X(0)=1$ and $dX(0)/du=0$, together with the Friedmann equation that describes the evolution of the scale factor $a(u)$. The numerical solution is evaluated up to a finite value of $u=u_{\rm end} = 20$ and extrapolated until the present time $\tau_0$ by matching it to the WKB solution $X \propto a^{-1}e^{\pm iu}$, which is substituted to Eq.~\eqref{eq:transfer_function}. We confirmed that the value of $\mathcal{T}_0h^2$ can be estimated with an accuracy of $\lesssim 0.2\,\%$ in our numerical scheme.
After obtaining the transfer function, the spectrum of GWs can be estimated by multiplying it by the primordial tensor power spectrum as shown in Eq.~\eqref{Omega_gw_def}.

In Fig.~\ref{fig:gw_events}, we show the spectrum of GWs for parameters corresponding to the benchmark point 1 in Table~\ref{tab:BMP_parameters}. As shown in this figure, various cosmological events occurring at temperature $T=T_{\rm hc}$ are imprinted on the shape of the GW spectrum at the corresponding frequency (cf. Eq.~\eqref{f_to_Thc}).
In this plot, the feature due to the PQ phase transition, which we are interested in, is less clear compared to that from $e^+e^-$ annihilation and the QCD crossover, since the change in the relativistic degrees of freedom is milder than in the latter cases. However, we will see below that such a feature could be distinguishable if a GW detector has enough sensitivity.

\begin{figure}[h]
\centering
\includegraphics[width=1.0\linewidth]{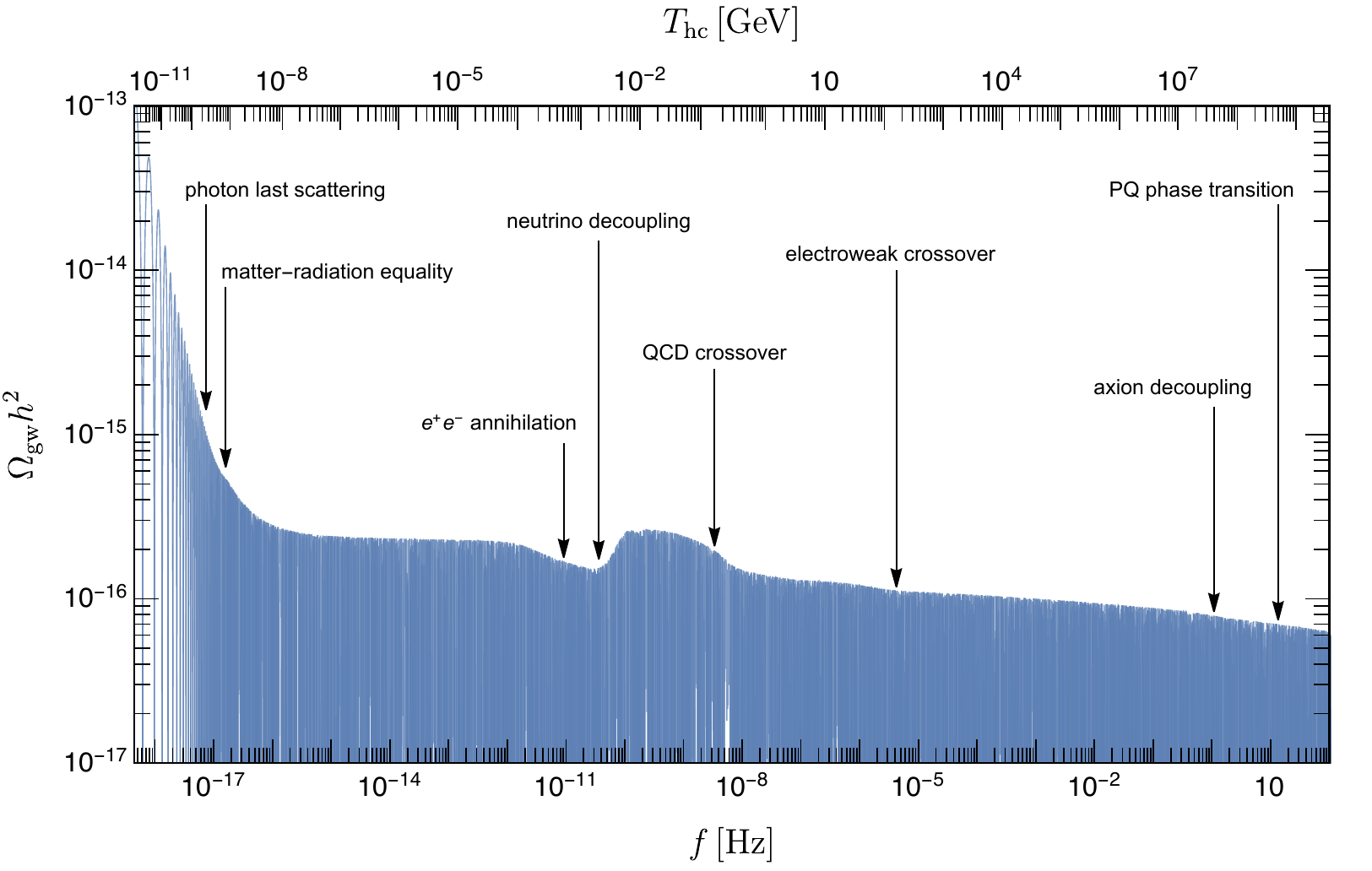}
\caption{The spectrum of GWs predicted in SMASH for a broad frequency interval.
The values of model parameters are fixed according to the benchmark point 1 in Table~\ref{tab:BMP_parameters}.
We also show the temperature $T_{\rm hc}$ corresponding to the horizon crossing of the mode with frequency $f$ evaluated according to Eq.~\eqref{f_to_Thc}.}
\label{fig:gw_events}
\end{figure}

The plot shown in Fig.~\ref{fig:gw_events} takes account of the fact that the amplitude of GWs coherently oscillate with a phase $2k\tau_0$. Practically, such oscillations cannot be resolved in direct detection experiments that are sensitive to GWs at high frequencies, and we may estimate the amplitude of GWs by replacing a rapidly oscillating factor by $1/2$. In the following, we use the notation $\Omega_{\rm gw}$ (or $\mathcal{T}_0$) to denote the averaged quantity.

As described in Eq.~\eqref{eq:int_diff_eq_for_chi}, we have included the damping effect due to free-streaming axions in our analysis of the GW spectrum. 
The coefficient $F_a(u)$ defined by Eq.~\eqref{eq:F_i_def} represents the efficiency of the damping effect, and it can be rewritten as
\begin{align}
F_a(u) = \frac{\frac{1}{2}\left(\frac{T_{\rm axion}}{T}\right)^4\left(\frac{g_{\ast s,0}}{g_{\ast s}(T)}\right)^{\frac{4}{3}}\Omega_{\gamma}h^2}{\Omega_Mh^2\left(\frac{a(u)}{a_0}\right)+\frac{g_{\ast\rho}(T)}{2}\left(\frac{g_{\ast s,0}}{g_{\ast s}(T)}\right)^{\frac{4}{3}}\Omega_{\gamma}h^2},
\label{eq:F_a_function}
\end{align}
where $\Omega_Mh^2 \simeq 0.14$ and $\Omega_{\gamma}h^2 \simeq 2.47 \times 10^{-5}$ are the present fraction of the energy densities of matter and photons, respectively.
In the radiation-dominated era, this function can be simplified as
\begin{align}
F_a(u) \to \frac{\left(T_{\rm axion}/T\right)^4}{g_{\ast\rho}(T)}.
\end{align}
Figure~\ref{fig:axion_damping} shows the impact of the axion damping effect on the spectrum of GWs.
Note that $F_a$ decreases monotonically for $T < T_{\rm dec}$, since $F_a = (T_{\rm axion}/T)^4 (g_{\ast\rho})^{-1} \propto (g^{\rm bath}_{\ast s})^{4/3}(g_{\ast\rho})^{-1} \propto (g^{\rm bath}_{\ast s})^{1/3}$ for $g^{\rm bath}_{\ast s} \simeq g_{\ast\rho}$.
On the other hand, $F_a$ loses its physical meaning for $T>T_{\rm dec}$ since axions do not exist as Goldstone bosons in this regime.
Therefore, the fraction of the energy density in relativistic axions is at most only $\sim 0.8\,\%$, which leads to a $\lesssim 1\,\%$ suppression of the amplitude of GWs 
at frequencies $f\lesssim 1\,\mathrm{Hz}$.
Although the effect is tiny, it might be relevant to future high-sensitivity GW experiments, as we discuss below.

\begin{figure}[h]
\centering
\includegraphics[width=0.5\linewidth]{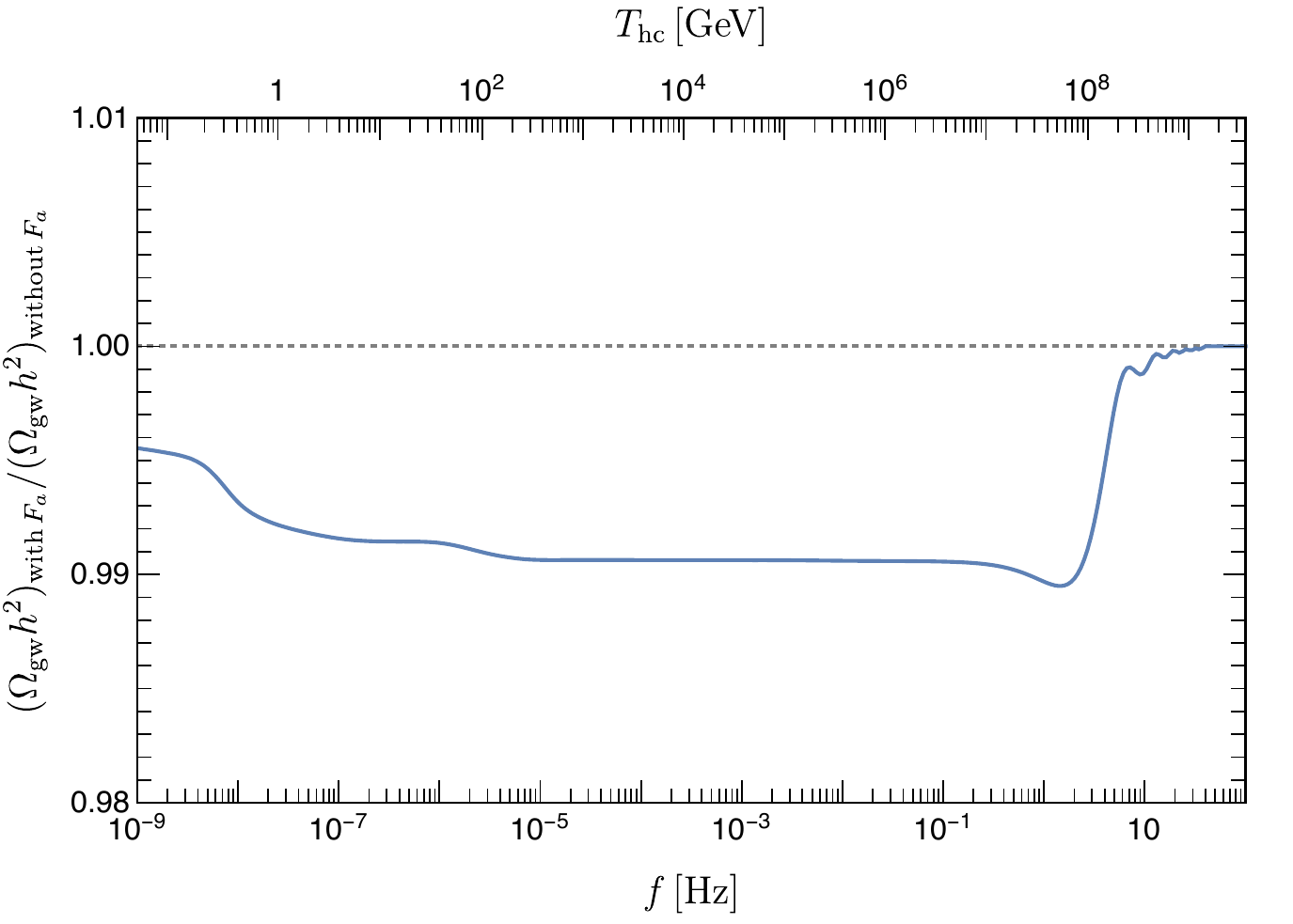}\includegraphics[width=0.5\linewidth]{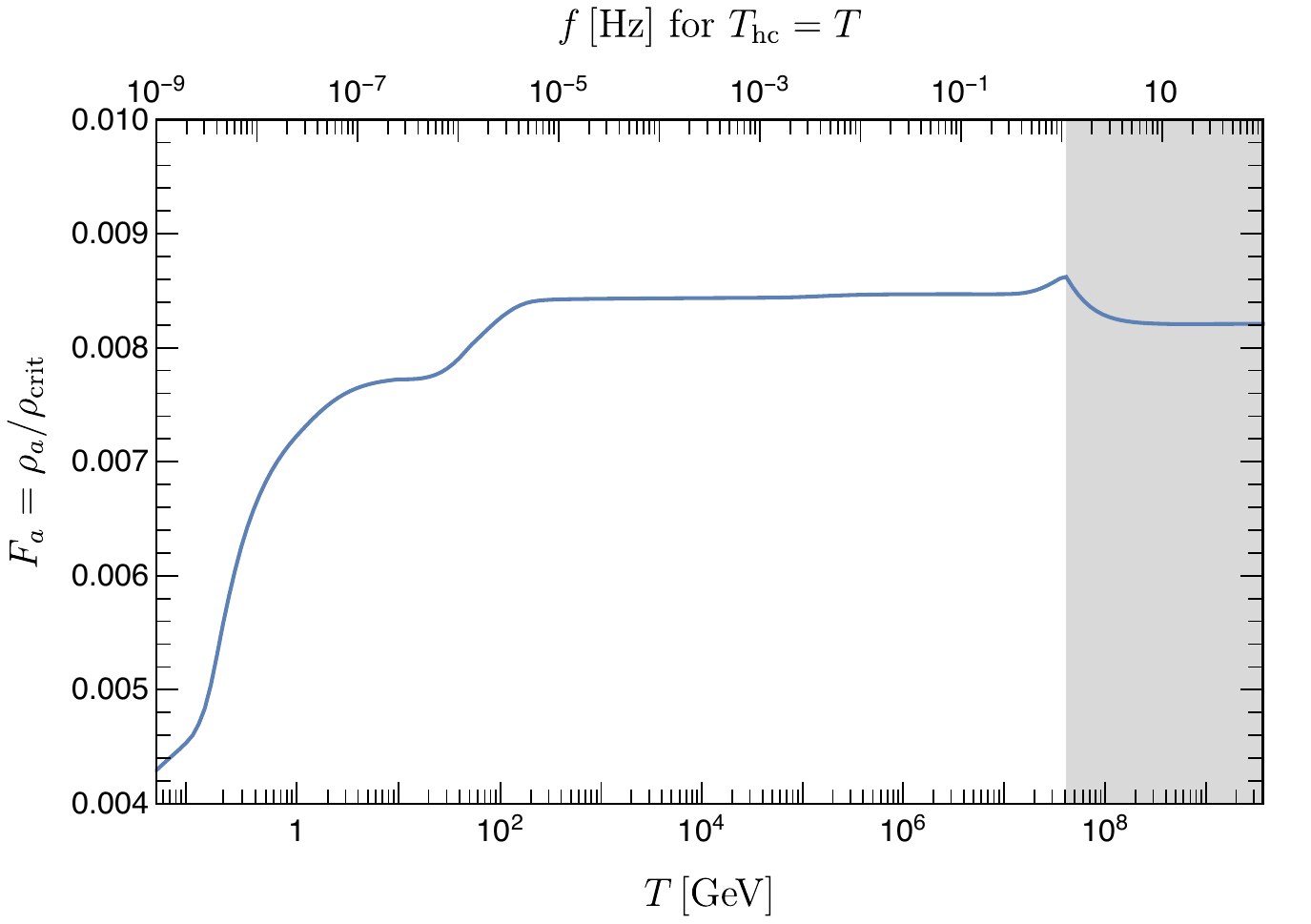}
\caption{Effect of free-streaming axions on the spectrum of GWs.
In the left panel, the ratio of $\Omega_{\rm gw}h^2$ obtained by including the axion damping effect to that obtained without including it (i.e. $F_a(u) = 0$) is plotted.
In the right panel, $F_a(u)$ given by Eq.~\eqref{eq:F_a_function} is plotted as a function of temperature $T$ or frequency $f$ that corresponds to $T_{\rm hc} = T$.
The shaded region in the right panel corresponds to $T > T_{\rm dec}$. 
These plots are produced based on the model parameters corresponding to the benchmark point 1 in Table~\ref{tab:BMP_parameters}.}
\label{fig:axion_damping}
\end{figure}

Figure~\ref{fig:transfer_function} shows the transfer function of GWs for the modes reentering the horizon around the epoch of the PQ phase transition.
It clearly shows that there is a step-like feature at $f \sim 1\,\mathrm{Hz}$ corresponding to the change in the effective relativistic degrees of freedom shown in Fig.~\ref{fig:eoszoom}.
In addition to this feature, the amplitude of GWs below that frequency is slightly suppressed due to the damping effect from free-streaming axions.
There is also a minor step-like feature at $f \sim 10^{-2}\,\mathrm{Hz}$, which is induced by the decoupling of $\rho$ particles.
These features are contrasted with the prediction in the SM, where the transfer function remains almost flat at these frequency ranges.\footnote{The transfer function in the SM
is not exactly flat and slightly red-tilted, because of the renormalization group running of gauge and Yukawa couplings~\cite{Saikawa:2018rcs}.}
The reason why the SMASH result with no axion damping (orange dashed line in Fig.~\ref{fig:transfer_function}) is larger than the SM (gray dotted line) at $f \lesssim 10^{-3}\,\mathrm{Hz}$
is just because the transfer function is proportional to $g_{\ast s,0}^{4/3}$ (see Eq.~\eqref{Omega_gw_est}) and its value in SMASH is larger than that in the SM, as shown in Eq.~\eqref{eq:gss0_SMASH}.

\begin{figure}[h]
\centering
\includegraphics[width=0.9\linewidth]{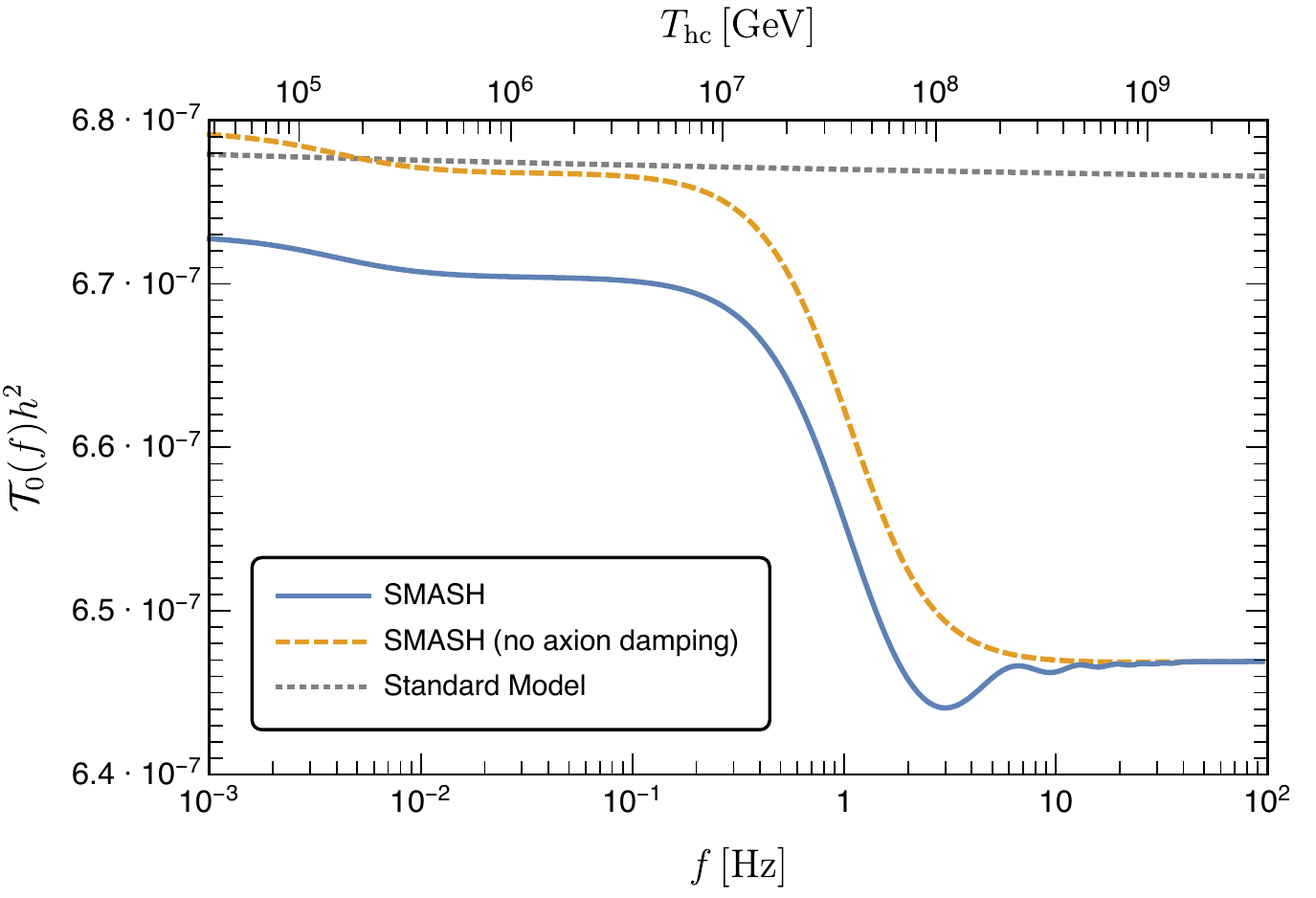}
\caption{
Frequency dependence of the transfer function $\mathcal{T}_0(f)h^2$ 
predicted in SMASH (blue solid line) and the SM (gray dotted line).
The result obtained by using the equation of state in SMASH, but not including the damping effect from free-streaming axions (i.e. $F_a(u) = 0$),
is also plotted as the orange dashed line for the sake of comparison.
The SMASH results shown in this figure are produced based on the model parameters corresponding to the benchmark point 1 in Table~\ref{tab:BMP_parameters}.}
\label{fig:transfer_function}
\end{figure}

The spectrum of GWs predicted in SMASH is compared with the projected sensitivities of future GW experiments in Fig.~\ref{fig:gwbroad_sensitivity}.
The experimental sensitivities in this figure represent the power-law integrated sensitivity (PLIS) curves~\cite{Thrane:2013oya} with a detection threshold 
given by the signal-to-noise ratio (SNR) of $\mathrm{SNR} = 1$,
which implies that any power-law signal that comes above these curves results in $\mathrm{SNR} > 1$.
The ongoing ground-based experiments such as the Advanced Laser Interferometer Gravitational Wave Observatory (aLIGO)~\cite{TheLIGOScientific:2014jea},
Advanced Virgo (aVirgo)~\cite{TheVirgo:2014hva}, and the Kamioka Gravitational Wave Detector (KAGRA)~\cite{Akutsu:2018axf} 
may not be sensitive enough to probe the range of GW amplitudes shown in this figure,
while more advanced detectors such as the Cosmic Explorer (CE)~\cite{Evans:2016mbw,Reitze:2019iox} 
and the Einstein Telescope (ET)~\cite{Punturo:2010zz} will improve the sensitivities at $f \sim \mathcal{O}(1\text{--}10)\,\mathrm{Hz}$.
For lower frequencies, the planned space-borne experiment LISA~\cite{Audley:2017drz,Baker:2019nia} is expected to probe GWs in the range $f \gtrsim 1\,\mathrm{mHz}$,
and future pulser timing array experiments, the International Pulsar Timing Array (IPTA)~\cite{Hobbs:2009yy} and the Square Kilometre Array (SKA)~\cite{Janssen:2014dka,Bull:2018lat}
will be sensitive to GWs at $f \sim 10^{-9}\,\mathrm{Hz}$.
Unfortunately, the nearly scale-invariant GW spectra predicted in the standard slow-roll inflationary models like SMASH are beyond the reach of the experiments mentioned above,
but there is a possibility that they can be probed by more advanced space-borne detectors like  BBO~\cite{BBO_proposal,Crowder:2005nr,Corbin:2005ny,Harry:2006fi} and DECIGO~\cite{Seto:2001qf,Kawamura:2006up}.
These experiments are designed to probe GWs at the frequency range $f\gtrsim 0.1\,\mathrm{Hz}$, which avoids the expected astrophysical foregrounds coming from 
extragalactic WD binaries~\cite{Farmer:2003pa}.\footnote{\label{fn:comments_WD}
In order to make a fair comparison between the PLIS curves and WD confusion noise,
we have used the following equation to draw the WD curve in Fig.~\ref{fig:gwbroad_sensitivity},
\begin{align}
\Omega_{\rm WD}(f) = \frac{4\pi^2}{3 H_0^2}\frac{f^3}{\sqrt{2 \Delta t_{\rm obs} \Delta f}}\frac{S_{\rm WD}(f)}{\Gamma(f)},
\label{eq:Omega_WD}
\end{align}
where $\Delta t_{\rm obs}$ is the observing time, $\Delta f$ is a range of frequencies at which the integrand giving the SNR becomes approximately constant,
 $S_{\rm WD}(f)$ is the strain noise power spectrum for the WD foregrounds, and $\Gamma(f)$ is the overlap reduction function of detectors.
Equation~\eqref{eq:Omega_WD} may represent a typical amplitude of the WD confusion noise in the correlation analysis between two GW detectors.
Here we take $\Delta t_{\rm obs} = 1\,\mathrm{year}$, $\Delta f = f/10$, and $\Gamma(f)=3/5$, assuming observations in LISA/DECIGO-like detectors,
and adopt the fitting formula for $S_{\rm WD}(f)$ taken from Ref.~\cite{Nishizawa:2011eq}.}
It is notable that the GW signals predicted by some of the benchmark points in SMASH are accessible to DECIGO and BBO.
On the other hand, in order to probe the detailed features on the GW spectrum, further improvement in the experimental sensitivities is indispensable.
This may be achievable in an idealized experiment whose sensitivity is limited by the quantum noise (ultimate DECIGO)~\cite{Seto:2001qf}.
Intriguingly, the frequency range $f\sim 1\,\mathrm{Hz}$ corresponding to the best sensitivity range of ultimate DECIGO
almost exactly coincides with the range where the non-trivial feature due to the PQ phase transition in SMASH emerges.

\begin{figure}[h]
\centering
\includegraphics[width=1.0\linewidth]{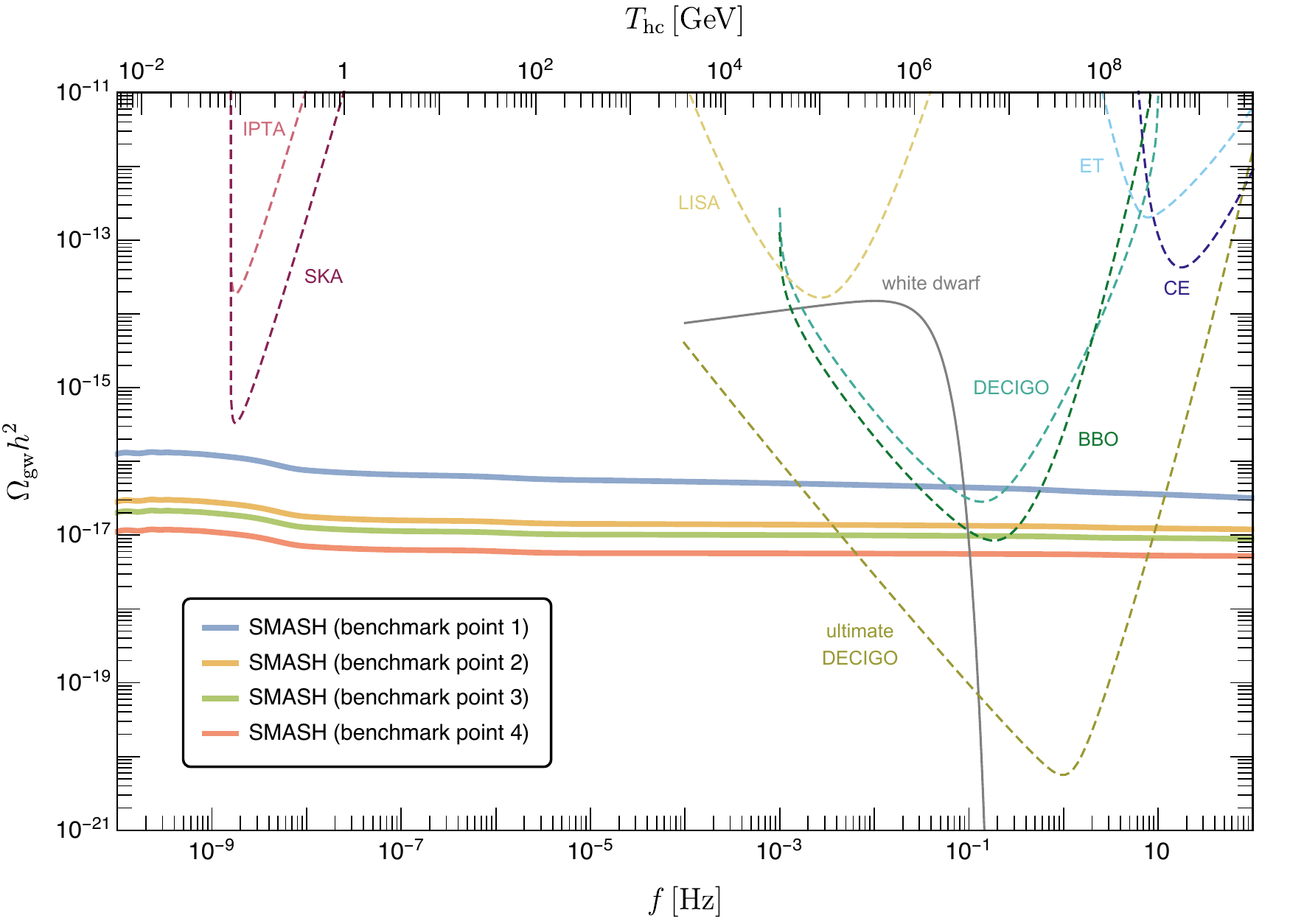}
\caption{
Spectrum of GWs predicted by the chosen SMASH benchmark points shown in Table~\ref{tab:BMP_parameters}
(thick solid lines) and projected sensitivities of future GW experiments represented as PLIS curves with $\mathrm{SNR} = 1$ (dashed lines).
All the sensitivity curves other than ultimate DECIGO are taken from Ref.~\cite{Schmitz:Zenodo} (see also Ref.~\cite{Schmitz:2020syl}).
The sensitivity curve for ultimate DECIGO is estimated by using the parameters specified in Ref.~\cite{Kuroyanagi:2014qza}.
An estimate of the astrophysical foregrounds from extragalactic WD binaries is also shown as gray thin sold line
(see footnote~\ref{fn:comments_WD} for details).}
\label{fig:gwbroad_sensitivity}
\end{figure}

More detailed comparisons between the GW signal in SMASH and the sensitivity of ultimate DECIGO are shown in Fig.~\ref{fig:signal}.
In this figure, we have estimated the uncertainty of $\Omega_{\rm gw}$ in ultimate DECIGO in terms of a discrete set of frequency bins,
following the procedure described in Appendix D of Ref.~\cite{Saikawa:2018rcs}.\footnote{The estimation of the sensitivity of ultimate DECIGO in this paper
is the same as Ref.~\cite{Saikawa:2018rcs} except for the following two modifications:
First, we take the angular efficiency factor (or the overall normalization of the overlap reduction function) as $F = 3/5$, which corresponds to the assumption that
the detectors consist of interferometers with opening angle of $\pi/3$ and make use of two independent data streams~\cite{Schmitz:2020syl},
rather than $F = 2/5$, which was used in Ref.~\cite{Saikawa:2018rcs} but corresponds to the case of interferometers with perpendicular arms and a single data channel.
Second, instead of assuming the constant overlap reduction function (i.e. $\Gamma(f) = \text{const.}$) 
at all frequencies, we include the decay of $\Gamma(f)$ at higher frequencies according to Fig.~10 of Ref.~\cite{Schmitz:2020syl}.
The first modification leads to an overall improvement of the sensitivity in $\Omega_{\rm gw}$ by a factor of $2/3$, while the second one
reduces the sensitivity at $f\gtrsim 10\,\mathrm{Hz}$.}
If the amplitude of the primordial tensor perturbation is sufficiently large, which is the case for the benchmark point 1 (top left panel of Fig.~\ref{fig:signal}), 
the primordial GW spectrum is more tilted because of the slow-roll consistency relation (see Eq.~\eqref{eq:n_t_consistency_relation}), 
and the feature due to the change in the effective relativistic degrees of freedom in SMASH is seen as an additional suppression of the tilted spectrum.
In this case, the shape of the GW spectrum can be identified over a wide frequency range, since the amplitude of GW itself is large
compared to the sensitivity of ultimate DECIGO.
On the other hand, the step-like feature becomes more manifest in the cases with smaller values of $r$, but 
in such cases the experimental noise is more severe because of the reduction of the overall GW amplitude.
The worst case is shown in the bottom right panel of Fig.~\ref{fig:signal} that corresponds to the benchmark point 4.
In spite of this difficulty, it is clear that in all four cases
the GW signatures resulting from the transfer function in SMASH are distinguishable from those in the SM,
which opens up the possibility of probing the non-trivial thermal history predicted in SMASH at future high-sensitivity GW experiments.

\begin{figure}[h]
\centering
\includegraphics[width=0.5\linewidth]{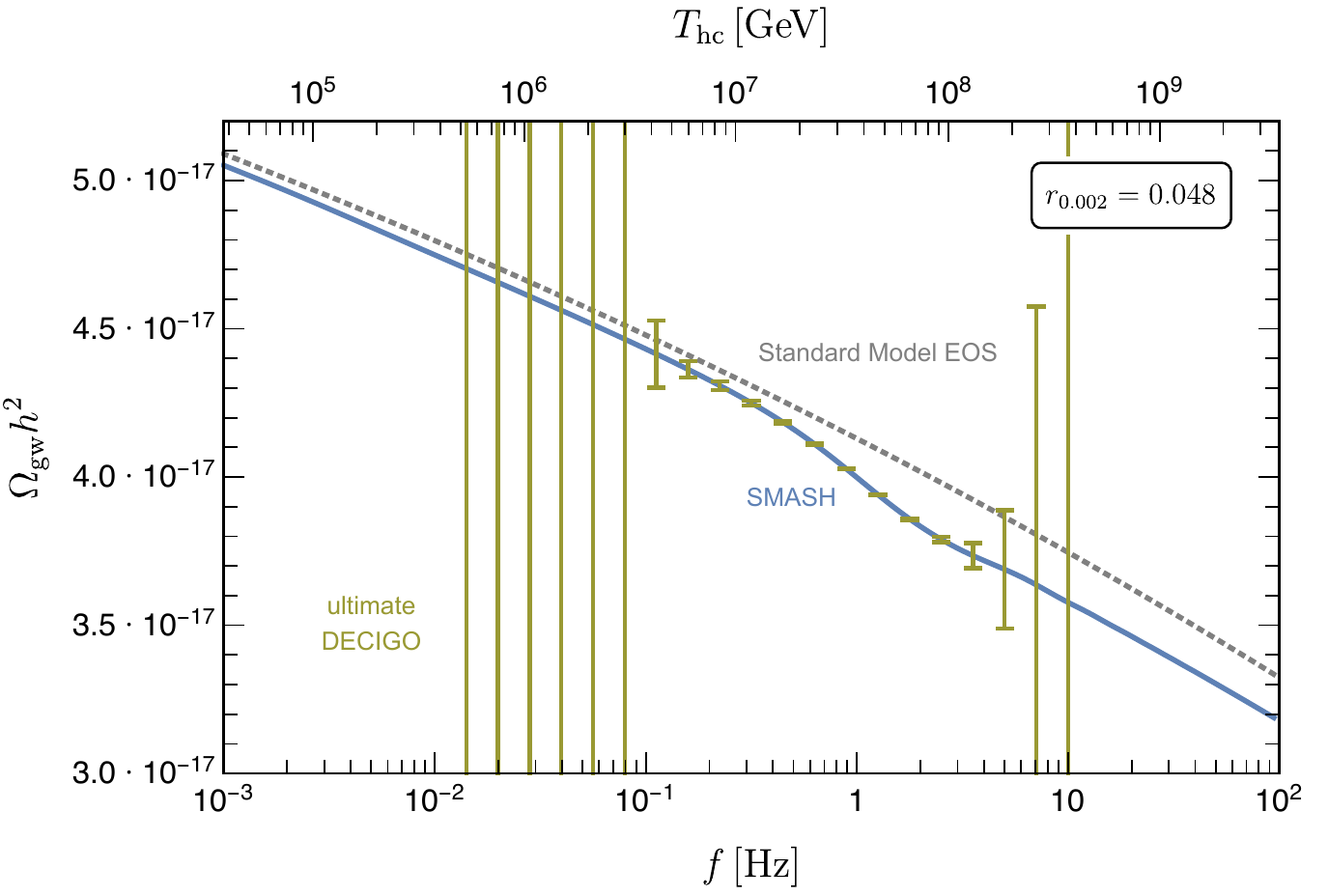}\includegraphics[width=0.5\linewidth]{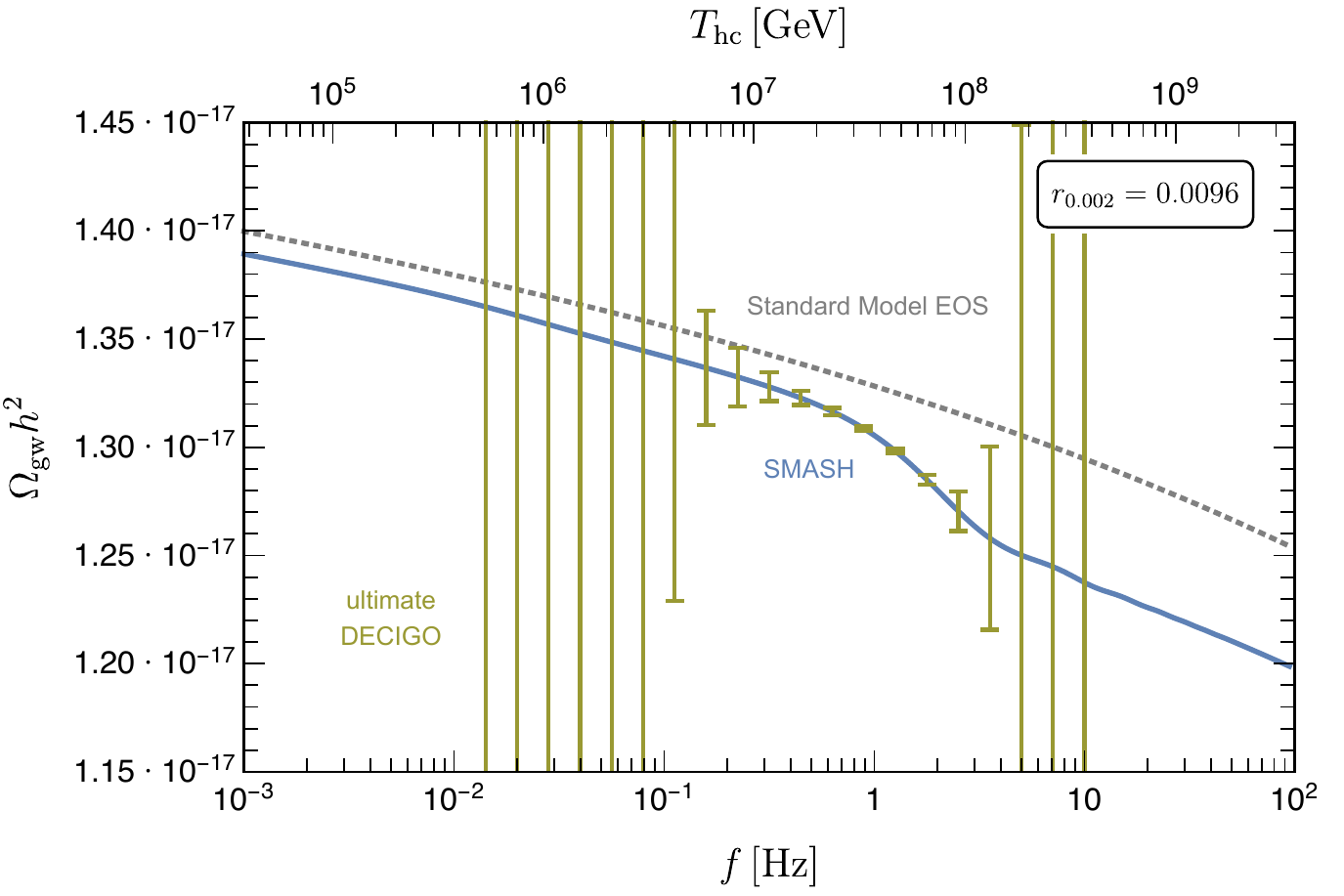}\\
\vspace{2mm}
\includegraphics[width=0.5\linewidth]{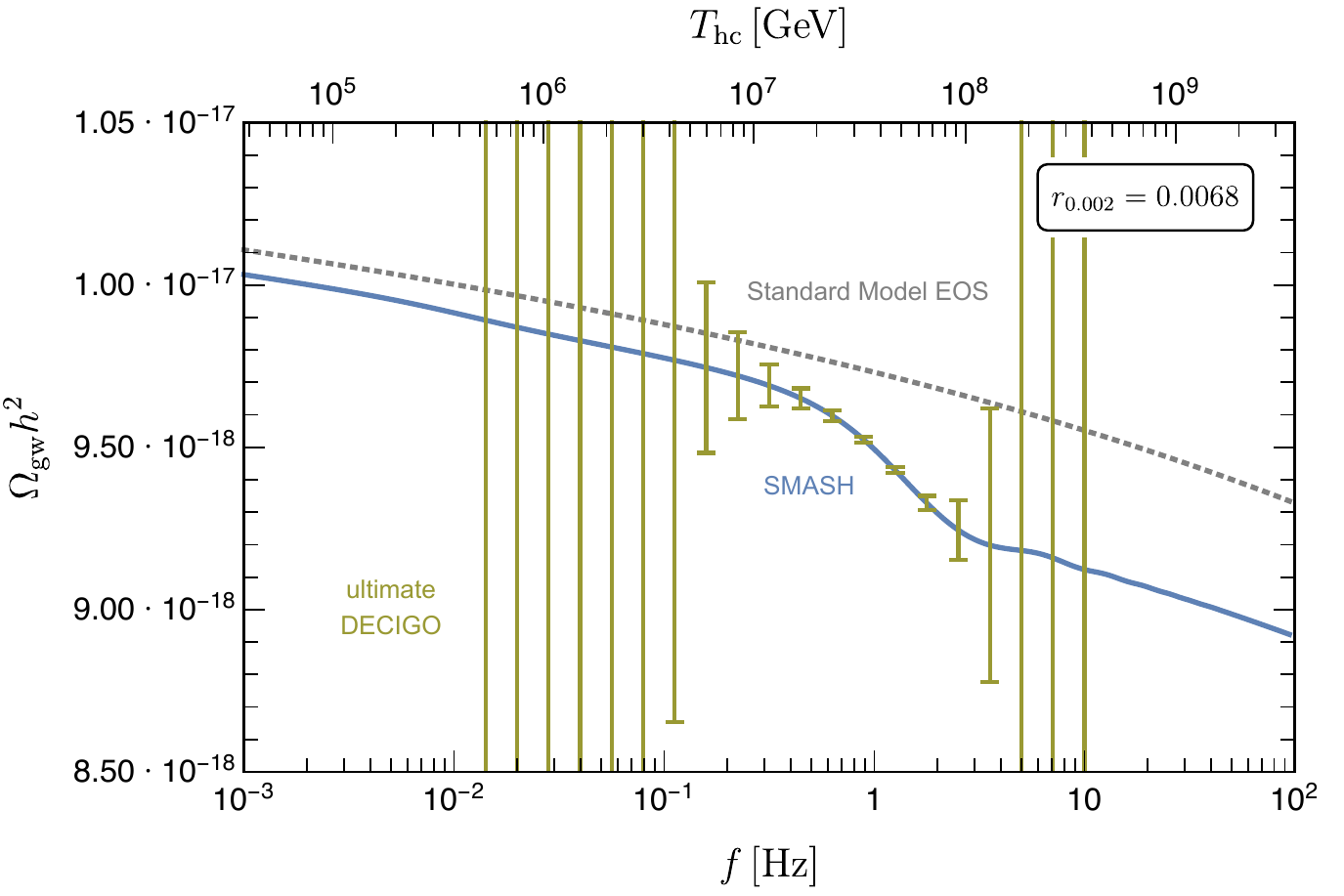}\includegraphics[width=0.5\linewidth]{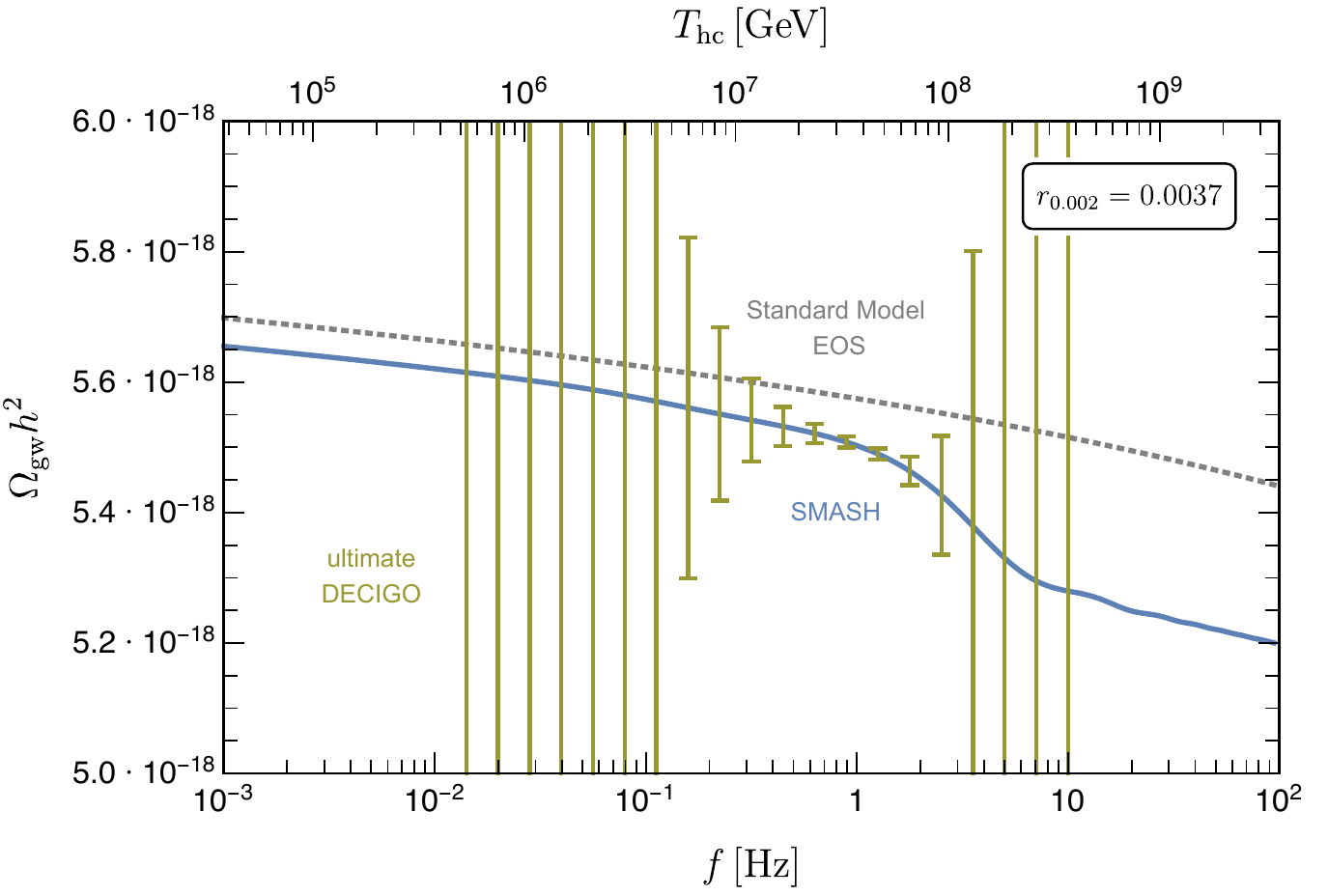}
\caption{The spectrum of primordial GWs predicted in SMASH benchmark point 1 (top left), 2 (top right),
3 (bottom left), and 4 (bottom right), corresponding to the choices of parameter values shown in Table~\ref{tab:BMP_parameters}.
Gray dotted lines represent the results obtained by using the effective relativistic degrees of freedom in the SM.
Dark yellow bars correspond to the sensitivity of ultimate DECIGO.}
\label{fig:signal}
\end{figure}

\section{Discussion and conclusions}
\label{sec:conclusions}
\setcounter{equation}{0}

In this paper we have focused on the primordial GWs predicted in SMASH. We have worked out precisely the unique prediction for the 
shape of the spectrum of GWs due to the nontrivial feature of the PQ phase transition described in this model.
The thermodynamic quantities in SMASH across the second order PQ phase transition were computed 
for four different benchmark parameter values shown in Table~\ref{tab:BMP_parameters} by adopting an improved daisy resummation scheme compatible with thermal decoupling,
which allowed us to estimate the effective relativistic degrees of freedom $g_{\ast\rho}(T)$ and $g_{\ast s}(T)$ for a wide temperature interval, as shown in Fig.~\ref{fig:eosfull}.
Using these estimates of $g_{\ast\rho}(T)$ and $g_{\ast s}(T)$, we computed numerically the spectrum of the primordial GWs 
originating from tensor fluctuations during inflation
and compared it to the sensitivity of ultimate DECIGO.
It turned out that the change in the relativistic degrees of freedom caused by the PQ phase transition in SMASH results in
a step-like feature in the spectrum of primordial GWs, and that ultimate DECIGO is capable to identify such a signal as shown in Fig.~\ref{fig:signal}.

As shown in Eq.~\eqref{f_to_Thc}, the modifications in the spectrum of the primordial GWs at particular frequency ranges can be related to cosmological events occurring at corresponding temperatures.
In this sense, the GW spectrum represents not only a snapshot of the early universe but also a movie looking deep into the very early universe, cf. Fig.~\ref{fig:gw_events}.
Different GW detectors sensitive to different frequency ranges can give snapshots of different epochs, 
providing us with a possibility to reconstruct this ``cosmic movie" by combining the results of various GW observations.
The SMASH model can be seen as a benchmark model for such a strategy, as it is a minimal model of particle cosmology,
allowing us to describe the whole thermal history of the universe from inflation until today in a decisive way as explicitly worked out in this paper.
Furthermore, one can also think of other minimal models of particle cosmology, such as for example the 
Neutrino Minimal SM ($\nu$MSM)~\cite{Asaka:2005an,Asaka:2005pn},  
a model based on a $B-L$ symmetry~\cite{Buchmuller:2012wn}\footnote{Attempts to
estimate the spectrum of GWs predicted in the model proposed in Ref.~\cite{Buchmuller:2012wn} were made in Refs.~\cite{Buchmuller:2013dja,Buchmuller:2019gfy}.}, and a model based on a flavour-dependent global $U(1)$ symmetry~\cite{Ema:2016ops}, and study the spectrum of GWs predicted in such models. 
The formalism developed in this paper may also be used for such endeavors.

Given that real direct observations of the primordial GWs in far advanced detectors like ultimate DECIGO will become feasible within this century~\cite{Seto:2001qf},
an important and necessary step in the near future is that the next generation CMB experiments should find primordial B-mode polarization patterns and measure 
an exact value of the tensor-to-scalar ratio $r$ at the CMB scale.
If it will be the case, we will be able to resolve the uncertainty regarding the value of $r$ (and hence the value of the effective inflationary coupling $\tilde{\lambda}_{\sigma}$)
and make more decisive predictions for the spectrum of GWs in SMASH at scales relevant to the space-borne GW experiments.
Then, at the time when such direct detection experiments start operating, one can obtain a richer information about the shape of the spectrum of primordial GWs and 
hence about viable parameter values in SMASH, by combining the results of CMB and direct detection experiments.
Such joint measurements of the GW spectrum over broad frequency ranges in various different detectors were considered recently in Ref.~\cite{Campeti:2020xwn}.

Throughout this paper, we have assumed a certain value for the scale of the PQ symmetry breaking or axion decay constant,
$v_{\sigma} = f_a = 1.2\times 10^{11}\,\mathrm{GeV}$.
In axion models with the post-inflationary PQ symmetry breaking and domain wall number $N_{\rm DW} =1$ (including SMASH), 
the value of $f_a$ should be determined from the requirement that the axion becomes the main constituent of dark matter.
Unfortunately, the determination of $f_a$ based on this cosmological argument currently involves a large uncertainty due to the fact that
the calculation of the axion dark matter abundance from global strings remains far from straightforward, 
despite a lot of effort made by several groups~\cite{Hiramatsu:2010yu,Hiramatsu:2012gg,Kawasaki:2014sqa,Klaer:2017qhr,Klaer:2017ond,Gorghetto:2018myk,Vaquero:2018tib,Buschmann:2019icd,Hindmarsh:2019csc,Drew:2019mzc}
(see the third bullet point between Eqs.~\eqref{eq:T_c_analytical} and~\eqref{eq:T_c_estimate}).
The latest results of numerical simulations of axionic strings reported in Ref.~\cite{Gorghetto:2020qws}, which take account of not only the radiation of axions in the scaling regime
but also the dynamics at the nonlinear regime when the effect of the axion mass becomes relevant,
imply that $f_a \lesssim 10^{10}\,\mathrm{GeV}$ when one extrapolates the simulation results to the realistic value of the string tension.
On the other hand, the analysis in Ref.~\cite{Klaer:2017ond} based on the effective action method that directly simulates strings with large tension~\cite{Klaer:2017qhr} 
gives a larger value, $f_a = (2.21\pm 0.29)\times 10^{11}\,\mathrm{GeV}$.
A careful assessment to understand the source of discrepancy between these results is still missing at the moment.

The determination of the value of $f_a$ is crucial for the calculation of the spectrum of the primordial GWs in SMASH,
since it affects the critical temperature of the PQ phase transition (see Eqs.~\eqref{eq:T_c_analytical} and~\eqref{eq:T_c_estimate}) 
and changes the frequency range at which the nontrivial feature on the GW spectrum emerges.
We generically expect that the signal from the PQ phase transition may become more difficult to detect if the required value of $f_a$ becomes lower, 
since the feature arising from the change in the effective relativistic degrees of freedom
will shift to lower frequencies and be hidden by the WD foreground contamination noise appearing at $f\lesssim 0.1\,\mathrm{Hz}$.

Independently of the theoretical predictions of $f_a$ mentioned above, its value will be fixed definitely once the axion dark matter is
found in the future haloscope experiments, such as 
the Axion Dark Matter eXperiment (ADMX)~\cite{Du:2018uak}, 
Capp Ultra Low Temperature Axion Search in Korea (CULTASK)~\cite{Chung:2016ysi}, 
Haloscope at Yale Sensitive to Axion CDM Experiment (HAYSTAC)~\cite{Zhong:2018rsr}, 
MAgnetized Disc and Mirror Axion eXperiment (MADMAX)~\cite{TheMADMAXWorkingGroup:2016hpc,Brun:2019lyf}, 
and Relic Axion Detector Exploratory Setup (RADES)~\cite{Melcon:2018dba}.
It is likely that the axion dark matter would be detected already
at the epoch when the advanced GW detectors like ultimate DECIGO are taking data,
if it actually constitutes the main component of dark matter.
Hence, the measurements of the spectrum of GWs should bring us  some further information about the model,
such as the self coupling of the PQ field, the Higgs portal coupling,
and the number and masses of exotic fermions, in addition to the value of the PQ scale.
In this sense, the future GW observations can be used to probe the details of the PQ sector that may not be reached in other high-energy experiments.

\section*{Acknowledgments}
We would like to thank Valerie Domcke for discussions and useful comments.
AR acknowledges partial  funding by the Deutsche Forschungsgemeinschaft (DFG, German Research Foundation) under Germany's Excellence Strategy – EXC 2121 ``Quantum Universe". 
Part of this research was supported by the Munich Institute for Astro- and Particle Physics (MIAPP) which is funded by DFG
under Germany's Excellence Strategy -- EXC-2094 -- 390783311.
This work of KS is supported by Leading Initiative for Excellent Young Researchers (LEADER), the Ministry of Education, Culture, Sports, Science, and Technology (MEXT), Japan. CT acknowledges financial support by SFB 1258 of DFG.

\appendix

\section{\label{app:Daisy}Improved daisy resummation compatible with thermal decoupling}

As the values of $g_{\ast \rho}$ and $g_{\ast s}$ depend on temperature, an accurate calculation has to incorporate all relevant thermal effects. As is well-known from thermal field theory, thermal corrections for two-point functions can become of the same order as their zero-temperature counterparts. Indeed, in the high-temperature limit and at low momentum it can be argued that  thermal contributions to bosonic self-energies are of the order of $g_i^2 T^2/m^2_{\rm tree}$ times the zero-temperature result, where $g_i$
stands for couplings of the theory, and $m_{\rm tree}$ is a zero-temperature mass. Therefore, a resummation is needed for $g_i T\gtrsim m_{\rm tree}$. In particular, this becomes necessary for temperatures above and near the critical temperature of phase transitions, as the latter are associated with effective masses of the form 
\begin{align}
\label{eq:m2eff}
m^2_{X,\rm eff}(T)= m^2_{X,\rm tree}+\Sigma_X(p=0,T)\sim m^2_{X,\rm tree}+f_X(g_i) T^2(1+O(m^2_{X,\rm tree}/T^2)) \,,
\end{align}
becoming zero for some particular bosonic field $X$ with self-energy $\Sigma_X$.
The resummation is achieved by substituting two-point functions by their thermally corrected counterparts, which is equivalent to resumming one-loop self-energy insertions in so-called ``daisy'' diagrams, as illustrated in Fig.~\ref{fig:Daisy}. In the low momentum limit, the resummation of self-energy insertions can be implemented as a thermal shift of the masses of the different bosonic particles as
\begin{align}
 m^2_{X,\rm tree}\rightarrow m^2_{X,\rm tree}+\Sigma_X(p=0,T)= m^2_{X,\rm eff}(T).
\end{align}
\begin{figure}
\begin{center}
 \includegraphics{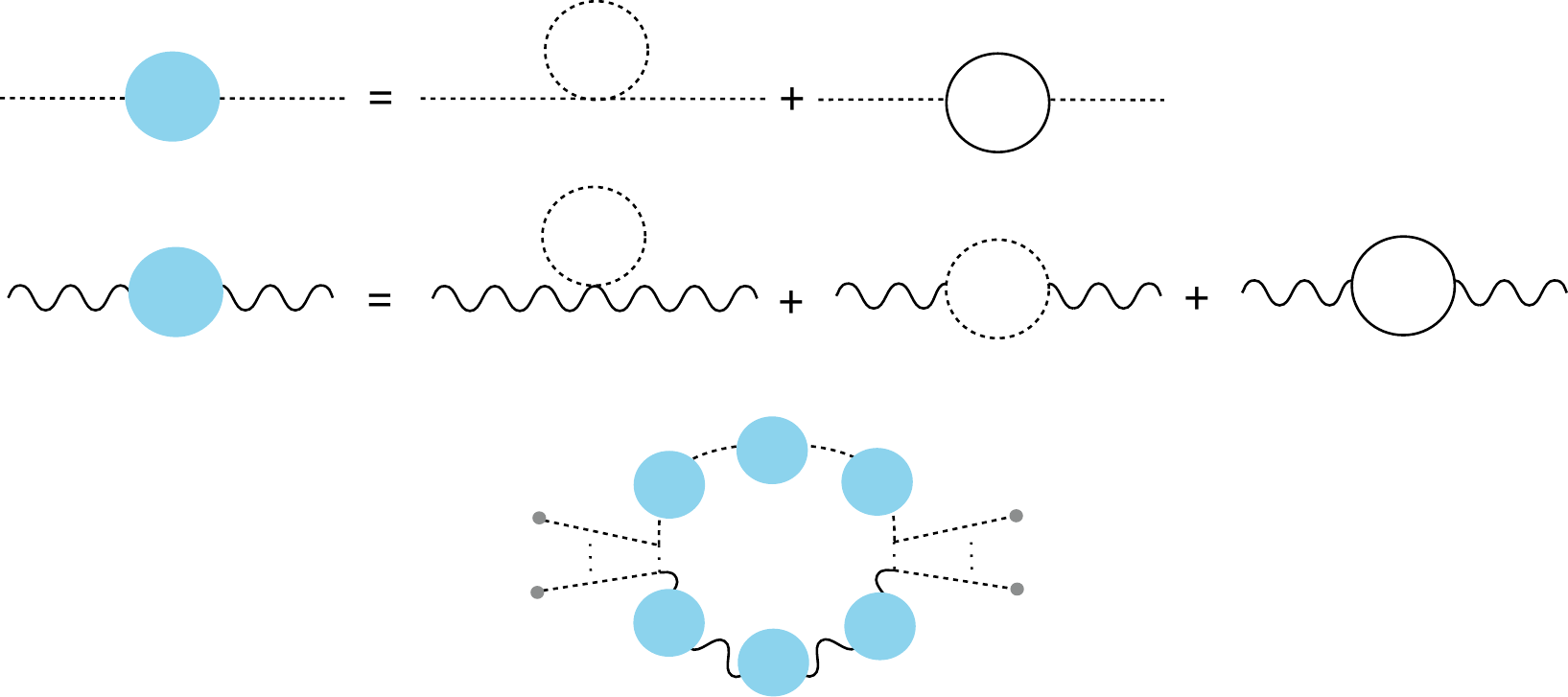}
\end{center}
 \caption{The upper two lines illustrate the corrections to the bosonic propagators coming from the scalars and fermions that become heavy during the PQ phase transition. The Daisy resummation in the effective potential accounts for all one-loop diagrams with external background legs and arbitrary insertions of the corrections to the internal two-point functions. This is illustrated at the bottom of the figure, where the dots in the external lines represent the background fields.}
\label{fig:Daisy} 
\end{figure}

Traditionally, one simply employs the leading corrections to $\Sigma_X(p=0,T)$ in a high-temperature expansion, as in Eq.~\eqref{eq:m2eff}. However, this approach is doomed to fail as the temperature goes down and the high-temperature expansion loses accuracy. This becomes crucially problematic for estimating the variation of the effective numbers of relativistic degrees of freedom, $g_{\ast\rho}$ and $g_{\ast s}$,  across a phase transition. The former quantities change during the transition precisely because some particles in the thermal plasma become nonrelativistic, such that $T/m_{X,\rm eff}$ drops below one and the particles decouple from the plasma. This decoupling is reflected by Boltzmann suppression factors in the corresponding contributions to the self-energies, which are not captured with the traditional daisy resummation. Hence, if one uses it to estimate $g_{\ast\rho}$ and $g_{\ast s}$ the results become unphysical.

Nevertheless, one can improve the usual daisy resummation by avoiding the high-$T$ expansion in the thermal corrections to the self-energies, while still keeping the zero-momentum approximation. As will be seen next, the resulting thermal corrections to the self-energies (which are absorbed into corrections to the masses) can be then expressed in terms of thermal integrals related to the finite-temperature corrections to the effective potential (e.g. the functions~\eqref{eq:Js} appearing in~\eqref{eq:VT}). We first consider the contribution to the thermal mass of a scalar singlet induced by other scalars and fermions, to be followed by the calculation of the contribution to the mass of an Abelian gauge boson induced by fermions. This covers the relevant cases in SMASH, where the fields becoming heavy during the phase transition are the heavy quarks and the real component of the complex scalar singlet. These particles contribute to the thermal masses of the Higgs boson, the hypercharge gauge boson, and the complex scalar of SMASH.

For the first case we assume a real scalar $\phi$ with a Yukawa interaction with a Dirac fermion $\psi$ and a quartic interaction with a real scalar $\chi$:
 \begin{align}\label{eq:Laux}
  {\cal L}\supset\left(-m_\psi\bar\psi\psi -\frac{\hat y}{\sqrt{2}} \phi \overline{\psi}P_L\psi+c.c.\right)-\frac{m^2_\phi}{2}\phi^2-\frac{m^2_\chi}{2}\chi^2-\frac{\lambda_\phi}{4!}\,\phi^4-\frac{\lambda_\chi}{4!}\,\chi^4-\frac{\lambda }{4}\,\phi^2 \chi^2.
 \end{align}
 For applications to SMASH, one can either interpret $\chi$ as a real component of the Higgs doublet and $\phi$ as a real component of the complex singlet, or the other way around; one simply has to use different mappings of the couplings in SMASH to the couplings in Eq.~\eqref{eq:Laux}.
 In the following, we will assume for simplicity that the Yukawa coupling $\hat{y}$ is real. For future reference, the thermal contributions to the effective potential of $\phi$ (we assume zero background for $\chi$) can be separated as:
 \begin{align} \begin{aligned}
  V^T=&\,V^{\phi,T}+V^{\chi,T}+V^{\psi,T},\\
  V^{\phi/\chi,T}=&\,\frac{T^4}{2\pi^2}\,J_B\left(\frac{m^2_{\phi/\chi}(\phi)}{T^2}\right),\\
   V^{\psi,T}=&\,-4\frac{T^4}{2\pi^2}\,J_F\left(\frac{m^2_{\psi}(\phi)}{T^2}\right).
 \end{aligned} \end{align}
 Note the extra factor of 2 in $V^{\psi,T}$, as we are dealing with a Dirac fermion composed of 2 Weyl fermions. The effective masses in a $\phi$ background are
 \begin{align} \begin{aligned}
  m^2_{\phi}(\phi)=&\,m^2_\phi+\frac{\lambda_\phi}{2}\phi^2,\\
  m^2_{\chi}(\phi)=&\,m^2_\chi+\frac{\lambda}{2}\phi^2,\\
  m_{\psi}(\phi)=&\,m_\psi+\frac{\hat y}{\sqrt{2}}\phi.
 \end{aligned} \end{align}

 We have the following one-loop corrections at finite $T$ for the $\phi$ self-energy in a $\phi$ background:
 \begin{align}\label{eq:Sigmas}\begin{aligned}
  -i\Sigma_\phi(0)=&\,-i\Sigma^\phi_\phi(0)-i\Sigma^\chi_\phi(0)-i\Sigma^\psi_\phi(0),\\
    -i\Sigma^{\phi}_\phi(0)=&\,\frac{i\lambda_\phi}{2\beta}\sum_{n=-\infty}^\infty\int \frac{d^3 l}{(2\pi)^3}\frac{1}{(2\pi n)^2-\vec{l}^2-m^2_{\chi}(\phi)},\\
  -i\Sigma^{\chi}_\phi(0)=&\,\frac{i\lambda}{2\beta}\sum_{n=-\infty}^\infty\int \frac{d^3 l}{(2\pi)^3}\frac{1}{(2\pi n)^2-\vec{l}^2-m^2_{\chi}(\phi)},\\
  -i\Sigma^\psi_\phi(0)=&\,-\frac{i \hat y^2}{2\beta}\sum_{n=-\infty}^\infty\int \frac{d^3 l}{(2\pi)^3}{\rm Tr}\,\frac{\slashed{l}\slashed{l}+O(m^2_{\psi}(\phi))}{\left((2\pi( n+1/2))^2-\vec{l}^2-m^2_{\psi}(\phi)\right)^2},\\
  =&\,-\frac{2i \hat y^2}{\beta}\sum_{n=-\infty}^\infty\int \frac{d^3 l}{(2\pi)^3}\frac{1}{(2\pi( n+1/2))^2-\vec{l}^2-m^2_{\psi}(\phi)}+O(\hat y^2m^2_{\psi}(\phi)),
 \end{aligned} \end{align}
 where we neglected contributions proportional $\hat y^2 m^2_\phi(\phi)$. The reason is that we focus on realizations with no tree-level mass terms for the fermions, so that $m^2_\phi(\phi)\propto \hat y^2$, and thus the terms  proportional to $\hat y^2 m^2_\phi(\phi)$ are $O(\hat y^4)$ and can be safely neglected for $\hat y\ll1$.
Using the identity (see e.g. \cite{Quiros:1999jp})
\begin{align}\label{eq:Trick}\begin{aligned}
\frac{1}{\beta} \sum_{n=-\infty}^{\infty} f(p^0=i\omega_n)=&\,\int_{-i\infty}^{i\infty}\frac{dz}{4\pi i}(f(z)+f(-z))+\eta_{F/B}\int_{-i\infty+\epsilon}^{i\infty+\epsilon}\frac{dz}{2\pi i} n_{F/B}(z)(f(z)+f(-z)),\\
\eta_F=&\,-1,\quad \eta_B=1,\quad n_{F/B}(z)=\frac{1}{e^{\beta z}-\eta_{F/B}},
\end{aligned}\end{align}
one can separate the loop integrals in Eq.~\eqref{eq:Sigmas} into zero $T$ and finite $T$ contributions, corresponding respectively to the first and second terms in Eq.~\eqref{eq:Trick}. Starting with the $\Sigma^{\chi}_\phi$ contribution, keeping only the finite-$T$ terms leads to
 \begin{align}
 \begin{aligned}
  -i\Sigma^{\chi,T}_\phi(0)=&\,\frac{\lambda}{2\pi}\int _{-i\infty+\epsilon}^{i\infty+\epsilon}dz\int\frac{ d^3 l}{(2\pi)^3}\frac{1}{e^{\beta z}-1}\cdot\frac{1}{z^2-\vec{l}^2-m^2_{\chi}(\phi)}.
 \end{aligned} \end{align}
 Closing the integration contour of $z$ on the ${\rm Re}\, z>0$ hemiplane, and picking up the residue at $z=E_l=\sqrt{\vec{l}^2+m^2_{\chi}(\phi)}$, with
 \begin{align}
  {\text{Res}}\,\left.\frac{1}{e^{\beta z}-1}\cdot\frac{1}{z^2+e^2_l}\right|_{z=E_l}=\frac{1}{2E_l}\frac{1}{e^{\beta E_l}-1},
 \end{align}
we get
\begin{align}
\begin{aligned}
  \Sigma^{\chi,T}_\phi(0)=&\,\frac{\lambda}{2}\int\frac{ d^3 l}{(2\pi)^3}\frac{1}{E_l}\frac{1}{e^{\beta E_l}-1}=\frac{\lambda}{4\pi^2}\int l^2 dl\frac{1}{\sqrt{l^2+m^2_{\chi}(\phi)}}\frac{1}{e^{\beta \sqrt{l^2+m^2_{\chi}(\phi)}}-1}.
 \end{aligned} \end{align}
 Changing variables to $y=l/T$, and defining $x\equiv m^2_{\chi}(\phi)/T^2$, we arrive at
\begin{align}
\begin{aligned}
  \Sigma^{\chi,T}_\phi(0)&\,=\frac{T^2\lambda}{4\pi^2}\int  dy\frac{y^2}{\sqrt{x+y^2}}\frac{1}{e^{\beta \sqrt{x+y^2}}-1}.
 \end{aligned} \end{align}
 From Eq.~\eqref{eq:Js} we see that this is related to derivatives of the thermal loop functions, and in fact to derivatives of the corresponding thermal contribution to the effective potential
 \begin{align}
 \begin{aligned}
  \Sigma^{\chi,T}_\phi(0)=&\,\frac{T^2\lambda}{2\pi^2}J_B'\left(\frac{m^2_{\chi}(\phi)}{T^2}\right)=\frac{T^2}{\pi^2}J_B'\left(\frac{m^2_{\chi}(\phi)}{T^2}\right)\frac{dm^2_{\chi}(\phi)}{d\phi^2}=2\frac{d}{d\phi^2}V^{\chi,T}.
 \end{aligned} \end{align}
 Of course, the last identity just confirms the fact that $\Sigma^{\chi,T}_\phi(0)$ can be interpreted as a contribution to the mass of $\phi$, since  the total mass goes $m^2\sim d^2V/(d^2\phi)|_{\phi=0}= 2 dV/(d\phi^2)|_{\phi=0}$.
 
 With the analytic expansion of $J_B$,
 \begin{align}
  J_B(x)=-\frac{\pi^4}{45}+\frac{\pi^2}{12}x+O(x^{3/2}),
 \end{align}
 we recover the shift of the mass of $\phi$ due to thermal $\chi$ loops in the high temperature limit,
 \begin{align}
  \Delta^\chi m^2_{\phi}= \Sigma_\phi^{\chi,T}(0)=\frac{\lambda}{24}\,T^2+\dots\,,
 \end{align}
 which is the result employed in traditional daisy resummations. As noted before, this is incompatible with decoupling, while the result that avoids the high $T$ expansion is general and perfectly compatible with decoupling:
  \begin{align}
  \Delta^\chi m^2_{\phi}= \frac{T^2}{\pi^2}J_B'\left(\frac{m^2_{\chi}(\phi)}{T^2}\right)\frac{dm^2_{\chi}(\phi)}{d\phi^2}.
 \end{align}
 The same reasoning applies to the contribution $\Sigma^{\phi,T}_\phi$. In the case of the fermions, repeating the same arguments as above, one has
 \begin{align}
   \Delta^\psi m^2_{\phi}= -\frac{4T^2}{\pi^2}J_F'\left(\frac{m^2_{\psi}(\phi)}{T^2}\right)\frac{dm^2_{\psi}(\phi)}{d\phi^2}=2\frac{d}{d\phi^2}V^{\psi,T}.
 \end{align}
 With the analytic expansion of $J_F$,
 \begin{align}
  J_F(x)=\frac{7\pi^4}{360}-\frac{\pi^2}{24}x+O(x^{3/2}),
 \end{align}
 one recovers the usual result for daisy resummations
 \begin{align}
   \Delta^\psi m^2_{\phi}= \frac{\hat y^2}{12}\,T^2+\dots.
 \end{align}

Finally, we can  consider the thermal mass of an Abelian boson $G$ (with associated gauge coupling $g$) induced by a heavy Dirac fermion $\psi$ with charge $q$. In the low momentum limit the thermal mass is captured by minus the temporal component of the self-energy \cite{Carrington:1991hz}, 
 \begin{align}
\Delta^\psi m^2_G=-\Pi_G^{\psi,00}(0)=
-\frac{q^2g^2}{\beta}\sum_{n=-\infty}^\infty\int \frac{d^3 l}{(2\pi)^3}\,{\rm Tr}\,\frac{(\slashed{l}+m_\psi)\gamma^0(\slashed{l}+m_\psi)\gamma^0}{\left((2\pi( n+1/2))^2-\vec{l}^2-m^2_{\psi}\right)^2}.
 \end{align}
 Proceeding in analogous manner as before the result is
 \begin{align}
  \Delta^\psi m^2_G=\frac{2g^2q^2T^2}{\pi^2} K_F\left(\frac{m^2_\psi}{T^2}\right),
 \end{align}
with $K_F$ given by Eq.~\eqref{eq:KF}. In the relativistic limit one has $K_F(0)={\pi^2/6}$, giving the standard daisy resummation
\begin{align}
 \Delta^\psi m^2_G=\frac{g^2q^2}{3}\,T^2+\dots.
\end{align}

\subsection*{Summary}

The previous formulae can be easily generalized to arbitrary backgrounds  involving more than one real scalar component, and for higher numbers of scalars and fermions. Labelling real scalars as $\phi_i$, and considering a background $\phi_i=\bar\phi_i$, the improved daisy resummation of self-energy corrections of a real scalar $\phi_j$ can be implemented by considering the following shift in the mass of $\phi_j$, arising from heavy real scalars $B=\{\phi_i\}$ and Weyl fermions $F$:

\begin{align}\label{eq:Deltam2phi}
 \Delta m^2_{\phi_j}(\bar\phi_i,T)\supset \sum_{B}\,\frac{T^2}{\pi^2}J_B'\left(\frac{m^2_{B}(\bar\phi_i)}{T^2}\right)\left.\frac{\partial m^2_{B}(\phi_i)}{\partial\phi_j^2}\right|_{\phi_i\rightarrow\bar\phi_i}-2\sum_{F}\,\frac{T^2}{\pi^2}\,J_F'\left(\frac{m^2_{F}(\bar\phi_i)}{T^2}\right)\left.\frac{\partial m^2_{F}(\phi_i)}{\partial \phi_j^2}\right|_{\phi_i\rightarrow\bar\phi_i}.
\end{align}
The derivatives of the thermal loop functions implement decoupling, as they go to zero exponentially as $m^2_X/T\gg1$. One can use numerical interpolating functions in the code. For the fields coupling to $\phi$ and remaining light (e.g. additional gauge bosons, ignored in the above formula), one can use the shift in the mass arising in the standard daisy resummation.

For the improved daisy resummation of the self-energy of an Abelian gauge field coupling to heavy Weyl fermions $F$, one can instead use the following shift in the mass:
\begin{align}\label{eq:Deltam2G}
  \Delta m^2_G\supset\sum_F\frac{g^2q^2T^2}{\pi^2} K_F\left(\frac{m^2_G(\bar\phi_i)}{T^2}\right).
\end{align}
The function $K_F$ becomes again exponentially suppressed if the argument is large, as expected from decoupling.
For additional light fields coupling to the Abelian boson one can follow the standard treatment.


\end{document}